\documentclass[12pt]{article}
\usepackage{newinutile3}
\usepackage{amsmath,amssymb}
\usepackage{mathrsfs}
\usepackage{epsfig,graphics,color,calc}
\usepackage{cite}
\usepackage{rotating}

\newtheorem{theorem}{Theorem}

\newtheorem{lemma}[theorem]{Lemma}

\newtheorem{proposition}[theorem]{Proposition}

\newtheorem{definition}[theorem]{Definition}

\newtheorem{condition}[theorem]{Condition}

\newcommand{\newatop}[2]{\genfrac{}{}{0pt}{}{#1}{#2}}
\newcommand{\qed}{$\phantom.$\hfill $\Box$\bigskip}
\newcommand{\rr}[1]{{\normalfont\textrm{#1}}}
\newcommand{\cc}[1]{{\mathcal{#1}}}
\newcommand{\bb}[1]{{\mathbb{#1}}}

\newcommand{\stab}{{x_0}}
\newcommand{\metauno}{{x_1}}
\newcommand{\metadue}{{x_2}}
\newcommand{\puno}{{\normalfont\textbf{u}}}
\newcommand{\muno}{{\normalfont\textbf{d}}}
\newcommand{\zero}{{\normalfont\textbf{0}}}
\newcommand{\pcri}{{\cc{P}_\mathrm{c}}}
\newcommand{\qcri}{{\cc{Q}_\mathrm{c}}}

\newcommand{\definisco}[1]{{\em{#1}}}

\newlength{\pecettawidth}
\def \ba {\begin{array}}
\def \ea {\end{array}}

\def \gate{{\mathcal{W}}}
\def \saddle{{\mathcal{S}}}

\def\1{\mathbbm{1}}

\begin{document}
\title{Sum of exit times in a series of two metastable states}

\author{Emilio N.M.\ Cirillo}
\affiliation{Dipartimento di Scienze di Base e Applicate per
             l'Ingegneria, Sapienza Universit\`a di Roma,
             via A.\ Scarpa 16, I--00161, Roma, Italy.}
\email{emilio.cirillo@uniroma1.it}

\author{Francesca R.\ Nardi}
\affiliation{Department of Mathematics and Computer Science,
             Eindhoven University of Technology,
             P.O.\ Box 513, 5600 MB Eindhoven, The Netherlands.}
\affiliation{Eurandom, P.O.\ Box 513, 5600 MB, Eindhoven, The Netherlands.}
\email{F.R.Nardi@tue.nl}

\author{Cristian Spitoni}
\affiliation{Department of Mathematics,
              Budapestlaan 6,
                  3584 CD Utrecht,The Netherlands}
\email{C.Spitoni@uu.nl}
\maketitle

\begin{abstract}
We consider the problem of non degenerate in energy metastable states
forming a series in the framework of reversible finite state space
Markov chains. We assume that starting from the
state at higher energy the system necessarily visits
the second one before reaching the stable state.
In this framework we give a sharp estimate of the exit time
from the metastable state at higher energy and, on the proper exponential
time scale, we prove an addition rule. As an application of the theory, we study the Blume--Capel model in the zero chemical potential case.

\end{abstract}




\section{Introduction}
\label{s:int}
\par\noindent
Metastable states in finite volume Statistical Mechanics lattice systems,
in which stochastic transition between states are controlled by an energy
function, is a well understood phenomenon.
Different mathematical theories have been developed in the last decades.
The pioneering \emph{pathwise approach} \cite{OS,MNOS,OV}, further developed for non--reversible dynamics in \cite{CNSo1},
and the
more recent \emph{potential theoretic approach}
\cite{BEGK,BdH,Slowick},
further developed in \cite{BL} via the use of the trace process,
allow a thorough description of the phenomenon.

The first theory gives a handy definition of the metastable
state and a physically clear interpretation of the
associated exit time. In the low temperature limit, 
it has been proven indeed that the
time scale on which the system leaves the metastable state
is exponentially large with the inverse temperature, with a mass
given by the smallest energy barrier that the system has to overcome
along the paths connecting the metastable state to the stable state. Moreover, the theory gives informations about
the paths followed by the system during its transition
to the stable state. In particular it ensures that with high
probability the system visits one of the configurations,
namely, the critical droplets, where the
smallest energy barrier is attained before reaching the stable state.

The latter theory, on the other hand, allows a precise estimate of the
mean exit time. In particular it is
proven that, measured in terms of the exponential time scale,
such a mean time can be 
written in terms of a quantity, called \emph{capacity}, which
can be bounded from above and from below by using effective
variational principles. Under suitable hypotheses on the
energy landscape, the theory allows the computation of the prefactor
and, what is very relevant on physical grounds, shows that it
does depend on entropic effects.
Indeed, in many specific models the prefactor has been computed explicitly
and it turns out to be connected, loosely speaking,
to the number of possible ways in which the system can perform its
transition to the stable state \cite{BM,MNOS}.
More precisely, it depends on the
number of critical droplets that the system can use as a gate
towards the stable state.

In this framework general results are proven under
suitable hypotheses on the energy landscape ruling out
the possibility to have multiple non degenerate in energy
metastable states. In this paper we approach precisely such a problem
and assume that two metastable states are indeed present.
Moreover, we chose a peculiar structure of the energy landscape
such that the two states form a ``series", in the sense that,
starting from the metastable state at higher energy, the system has
to visit necessarily (in probability in the low temperature limit)
the second one in its way towards the ground state.
We prove a sort of addition rule for the exit time and compute,
on the exponential scale, a sharp estimate for the exit time.

We approach this problem in the framework of rather general
reversible Markov chains. Our aim is that of introducing a model
covering both the standard Statistical Mechanics stochastic lattice
models (e.g., the Metropolis dynamics) and the
reversible Probabilistic Cellular Automata.
In this framework we assume a minimal characterization
of the energy landscape sufficient to ensure both the presence
of two non degenerate in energy metastable state
and their serial structure.

In the last part of the paper we discuss an application of
this theory. Indeed, we approach the Blume--Capel model \cite{B,C},
whose metastability behavior has already been studied from different
point of views and in different limits in
\cite{CO,MO,CN2013,LL2016},
and we derive, with a different method, the same result
recently proven in \cite{LL2016}
on the sharp estimate of the exit time from the
metastable minus one state in the zero chemical potential case.
We mention that the application of our results to the 
Probabilistic Cellular Automaton studied in \cite{CN}
is reported in \cite{CNSautomata2016}.

The paper is organized as follows: in Section~\ref{s:genris} the general
model is introduced and our main results are stated.
Those results are then proved in Section~\ref{s:dimo}.
Finally, the application to the Blume--Capel model is
discussed in Section~\ref{s:blume}.

\section{Model and results}
\label{s:genris}
\par\noindent
In this section we first introduce a general reversible Markov chain and
specify  the conditions on the energy landscape in order to have a series of
metastable states. We next state our main results in this framework.

\subsection{Reversible Markov chains}
\label{s:reversibile}
\par\noindent
We want to give the notion of reversible Markov chain
\cite[Condition R, page~335]{OV}
in a quite general setup
so that the theory
will apply to different
and relevant examples such as Statistical Mechanics Lattice models
and the reversible Probabilistic Cellular Automata.

Consider a finite state space $X$
and a family of irreducible and aperiodic Markov chain
$x(t)$, with $t\in\bb{Z}_+$
parametrized by the parameter
$\beta>0$, called \emph{inverse temperature}. We let
$p_\beta(x,y)$ and $\mu_\beta(x)$, for $x,y\in X$ be respectively
the transition matrix and the stationary measure.
We assume that the Markov chains are reversible with respect to
$\mu_\beta$, namely,
\begin{equation}
\label{dettagliato}
\mu_\beta(x)p_\beta(x,y)=p_\beta(y,x)\mu_\beta(y)
\end{equation}
for any $x,y\in X$.
We also assume that the fact that a jump is not allowed does not
depend on $\beta$, namely, if $p_{\beta'}(x,y)=0$ then
$p_{\beta''}(x,y)=0$ for any $\beta''$.

The definition of the model will be
completed by assuming a slightly enforced
version of the well known Wentzel--Friedlin condition and by requiring
that the stationary measure is ``close'' to a Gibbs one\footnote{The
fact the stationary measure is close to a Gibbs one at low temperature
is generally valid in the framework of Wentzell--Freidlin dynamics,
see, e.g., \cite[Proposition~4.1]{Ca}. What we assume in this paper,
see \eqref{rev04-gib}, is, on the other hand, slightly stronger.}.
More precisely, we assume that
there exists
$\Delta:X\times X\to\bb{R}_+\cup\{\infty\}$ and
$r:X\times X\to\bb{R}$
such that, for any $x,y\in X$, $\Delta(x,y)=\infty$ if 
$p_\beta(x,y)=0$
and
\begin{equation}
\label{rev04}
\lim_{\beta\to\infty} [-\log p_\beta(x,y)-\beta\Delta(x,y)]=r(x,y)
\;\;\textrm{ if } p_\beta(x,y)>0
.
\end{equation}
Note that $\Delta$ and $r$ do not depend on the inverse
temperature. We shall call $\Delta$
the \emph{cost function}.

Moreover, we assume that
there exist two functions
$s,H: X\to\bb{R}$ and a family of functions
$G_\beta:X\to\bb{R}$ parametrized by $\beta>0$ such that
\begin{equation}
\label{rev04-gib}
\mu_\beta(x)=\frac{1}{\sum_{y\in X}e^{-G_\beta(y)}}\,e^{-G_\beta(x)}
\;\;\textrm{ and }\;\;
\lim_{\beta\to\infty} [G_\beta(x))-\beta H(x)]=s(x)
\end{equation}
for $x\in X$ and $\max_{x\in X} s(x)=\bar{s}<\infty$.
The normalization factor in $\mu_\beta$ is denoted by $Z_\beta$ and
called \emph{partition function}.
Note that the function $H$ does not depend on the inverse
temperature. We shall call $H$
the \emph{Hamiltonian} or \emph{energy}
of the model.

From \eqref{rev04} and \eqref{rev04-gib}
it follows immediately that
\begin{displaymath}
\lim_{\beta\to\infty}\frac{1}{\beta}\log p_\beta(x,y)=-\Delta(x,y)
\end{displaymath}
for all $x,y\in X$  such that $p_\beta(x,y)>0$ and
\begin{displaymath}
\lim_{\beta\to\infty}\frac{1}{\beta}G_\beta(x)=H(x)
\end{displaymath}
for $x\in X$.
From \eqref{dettagliato} and the conditions above
it follows also that
\begin{equation}
\label{rev03}
H(x)+\Delta(x,y)=\Delta(y,x)+H(y)
\end{equation}
for all $x,y\in X$.

For any $x\in X$, we denote by
$\bb{P}_x(\cdot)$ and $\bb{E}_x[\cdot]$
respectively the probability and the average along the trajectories
of the process started at $x$.

\subsection{Examples}
\label{s:esempi}
\par\noindent
In this section we discuss two important examples of dynamics
fitting in the general scheme depicted above.
The \emph{Metropolis} dynamics on the states space $X$ with
\emph{energy} $K:X\to\bb{R}$, \emph{inverse temperature}
$\beta$, and \emph{connection matrix} $q(x,y)$  is defined by letting
\begin{displaymath}
p_\beta(x,y)
=
q(x,y)\,e^{-\beta[K(y)-K(x)]_+} \;\; \textrm{ if } x\neq y
\end{displaymath}
where, for any real $a$, we let $[a]_+=a$ if $a>0$ and
$[a]_+=0$ otherwise be the positive part of $a$,
and
\begin{displaymath}
p_\beta(x,x)
=
1-\sum_{y\neq x}p_\beta(x,y)
.
\end{displaymath}

It is well known that the Metropolis dynamics has as stationary measure
the Gibbs measure with Hamiltonian $K$, so that the condition
\eqref{rev04-gib} is satisfied with $G_\beta=\beta K$ and $s=0$.

Now, we let $r(x,x)=0$ and
$r(x,y)=-\log q(x,y)$ for $x\neq y$ and $q(x,y)>0$.
For any $x\neq y$,
we let also
$\Delta(x,y)=\infty$ if $q(x,y)=0$ and
$\Delta(x,y)=[K(y)-K(x)]_+$ otherwise.
Finally, we let
$\Delta(x,x)=\infty$ if $p_\beta(x,x)=0$ and
$\Delta(x,x)=-(1/\beta)\log p_\beta(x,x)$ otherwise.
It is immediate to verify that
condition \eqref{rev04} is satisfied for any $x,y\in X$.

A second important example is that of
reversible Probabilistic Cellular Automata (PCA).
Reversible PCA
have been introduced in \cite{GJH}, see also \cite{CLRS} for a detailed
discussion,
and provide a very interesting
example of dynamics with a parallel updating rule which are
reversible with respect to a stationary measure which is very
close to a Gibbs measure.
The metastable behavior of some reversible Probabilistic
Cellular Automata has been firstly studied in \cite{BCLS}.

Let $\Lambda\subset\bb{Z}^2$ be a finite cube with periodic boundary
conditions.
Associate with each site $i\in\Lambda$
the state variable
$x_i\in\{-1,+1\}$ and denote by
$X=\{-1,+1\}^\Lambda$ the \emph{state space}.
For any $x\in X$
we consider on $\{-1,+1\}$ the probability
measure
\begin{displaymath}
 f_{x,\beta}(s)=\frac{1}{2}\Big\{
        1+s\tanh\Big[\beta\Big(\sum_{j\in\Lambda}k(j)x_j+h\Big)\Big]\Big\}
\end{displaymath}
for $s\in\{-1,+1\}$,
where $\beta>0$
and $h\in\mathbb{R}$ are called \textit{inverse temperature} and
\textit{magnetic field} respectively.
The function
$k:\bb{Z}^2\to\bb{R}$ is such that its
support is a subset of~$\Lambda$ and~$k(j)=k(-j)$
for all  $j\in\Lambda$.
Recall that, by definition, the support of the function $k$ is the subset
of $\Lambda$ where the function~$k$ is different from zero.
We assume, also, that
\begin{equation}
\label{pca00}
\sum_{j\in\Lambda}k(j-i)x_j+h\neq 0
\end{equation}
for any $x\in X$ and $i\in\Lambda$.

We finally introduce the shift $\Theta_i$ on the torus, for any $i\in\Lambda$,
defined as the map
$\Theta_i:X\to X$ such that
$(\Theta_i x)_j=x_{i+j}$.
A reversible PCA is
the Markov chain
on $X$ with transition matrix
\begin{displaymath}
p_\beta(x,y)=\prod_{i\in\Lambda}f_{{\Theta_{i}x},\beta} (y_i)
\end{displaymath}
for $x,y\in X$.
We
remark that
the character of the
evolution is parallel, in the sense that at each time
all the spins are potentially flipped.

It is not difficult to prove~\cite{GJH}
that the above specified PCA dynamics is reversible
with respect to the finite--volume Gibbs--like
measure
\begin{displaymath}
 \mu_{\beta}(x)
 =
 \frac{1}{Z_\beta}
\,e^{-F_{\beta}(x)}
\end{displaymath}
with
\begin{displaymath}
F_{\beta}(x)
=
 -\beta h\sum_{i\in\Lambda}x_i
 -\sum_{i\in\Lambda}
    \log\cosh\Big[\beta
   \Big(\sum_{j\in\Lambda}k(j-i)x_j+h\Big)\Big]
\end{displaymath}
with $Z_\beta$ the normalization constant.

The
low--temperature behavior of the stationary measure of the PCA
can be guessed by looking at the function
\begin{displaymath}
K(x)
=
\lim_{\beta\to\infty}\frac{1}{\beta}F_{\beta}(x)
=
-h\sum_{i\in\Lambda}x_i
-\sum_{i\in\Lambda}
   \Big|\sum_{j\in\Lambda}k(j-i)x_j+h\Big|
\end{displaymath}
The difference between
$F_{\beta}$ and $\beta K$ can be computed explicitly,
indeed in \cite{CLRS} it is proven that
\begin{displaymath}
F_{\beta}(x)-\beta K(x)
=
-
\sum_{i\in\Lambda}
 \log\Big(
          1+\exp\Big\{
                      -2\beta
                        \Big|\sum_{j\in\Lambda}k(j-i)x_j+h\Big|
                \Big\}
     \Big)
+|\Lambda|\log2
\end{displaymath}
for each $\beta>0$ and $x\in X$.

From the remarks above, recall also the assumption
\eqref{pca00}, it follows immediately that the
reversible PCA satisfies condition \eqref{rev04-gib}
with $G_\beta=F_{\beta}$, $H=K$, and $s(x)=|\Lambda|\log 2$
for any $x\in X$.

As for the transition rates, we set
\begin{displaymath}
V(x,y)
=-\lim_{\beta\to\infty}\frac{1}{\beta}\log p_\beta(x,y)
=\hspace{-1.0cm}\sum_{\newatop{i\in\Lambda:}
               {y_i(\sum_{j\in\Lambda}k(j-i)x_j+h)<0}}
\hspace{-0.4cm}2\Big|\sum_{j\in\Lambda}k(j-i)x_j+h\Big|
\end{displaymath}
We shall prove that
\begin{equation}
\label{accaede}
-\log p_\beta(x,y)-\beta V(x,y)
=
 \sum_{i\in\Lambda}
 \log(1+e^{-2\beta|\sum_{j\in\Lambda }k (j-i)x_i+h|})
\end{equation}
Thus, the reversible PCA satisfies condition \eqref{rev04}
with $\Delta=V$ and $r(x,y)=0$
for any $x,y\in X$.

For completeness, we finally prove \eqref{accaede}. In the
following computation we shall use many times the assumption
\eqref{pca00}. First note that
\begin{displaymath}
\begin{array}{l}
{\displaystyle
 -\log p_\beta(x,y)-\beta V(x,y)=
\sum_{i\in\Lambda}
 \log(1+e^{-2\beta y_i[\sum_{j\in\Lambda }k (j-i)x_j+h]})
}
\\
{\displaystyle
\phantom{mermermermermermremreme}
\phantom{=}+
\beta
\!\!\!\!
\!\!\!\!
\sum_
{\newatop{i\in\Lambda:}{y_i(\sum_{j\in\Lambda}k(j-i)x_j+h)<0}}
\!\!\!\!
\!\!\!\!
   2y_i\Big(\sum_{j\in\Lambda}k(j-i)x_j+h\Big)
}
\end{array}
\end{displaymath}
Hence
\begin{displaymath}
\begin{array}{l}
{\displaystyle
 -\log p_\beta(x,y)-\beta V(x,y)=
\sum_{\newatop{i\in\Lambda:}{y_i(\sum_{j\in\Lambda}k(j-i)x_j+h)>0}}
\hspace{-1cm}
 \log(1+e^{-2\beta y_i[\sum_{j\in\Lambda }k (j-i)x_j+h]})
}
\\
{\displaystyle
\phantom{mermermermermermremreme}
+
\!\!\!\!
\!\!\!\!
\sum_{\newatop{i\in\Lambda:}{y_i(\sum_{j\in\Lambda}k(j-i)x_j+h)<0}}
\hspace{-1cm}
 \log(1+e^{-2\beta y_i[\sum_{j\in\Lambda }k (j-i)x_j+h]})
}
 \vphantom{\Bigg\{_\Bigg\}}
\\
{\displaystyle
\phantom{mermermermermermremreme}
+
 \beta
\!\!\!\!
\!\!\!\!
 \sum_
{\newatop{i\in\Lambda:}{y_i(\sum_{j\in\Lambda}k(j-i)x_j+h)<0}}
\!\!\!\!
\!\!\!\!
 2y_i\Big(\sum_{j\in\Lambda}k(j-i)x_j+h\Big)
}
 \vphantom{\Bigg\{_\Bigg\}}
\end{array}
\end{displaymath}
Finally, 
\begin{displaymath}
\begin{array}{l}
{\displaystyle
 -\log p_\beta(x,y)-\beta V(x,y)=
\sum_{\newatop{i\in\Lambda:}{y_i(\sum_{j\in\Lambda}k(j-i)x_j+h)>0}}
\hspace{-1cm}
 \log(1+e^{-2\beta y_i[\sum_{j\in\Lambda }k (j-i)x_j+h]})
}
\\
{\displaystyle
\phantom{mermermermermermremreme}
+
\!\!\!\!
\!\!\!\!
\sum_{\newatop{i\in\Lambda:}{y_i(\sum_{j\in\Lambda}k(j-i)x_j+h)<0}}
\hspace{-1cm}
 \log(e^{2\beta y_i[\sum_{j\in\Lambda }k (j-i)x_j+h]}+1)
}
\\
\end{array}
\end{displaymath}
yielding (\ref{accaede}).

\subsection{Energy landscape}
\label{s:rel}
\par\noindent
After the short ``intermezzo'' on the Metropolis and the reversible
PCA models, we come back to the general setup of Section~\ref{s:reversibile}.
Let $Q$ be the set of pairs $(x,y)\in X\times X$ such that
$p_\beta(x,y)>0$ or, equivalently, $\Delta(x,y)<\infty$.
The quadruple
$(X,Q,H,\Delta)$ is then a reversible energy landscape
\cite{CN2013}.

Given $Y\subset X$ we let its \textit{external boundary} $\partial Y$
be the collection of states $z\in X\setminus Y$ such that there exists
$y\in Y$ such that $(y,z)\in Q$. In words, the external boundary
is made of those states outside $Y$ such that there exists a state
in $Y$ where the system can jump.

Given $Y\subset X$ such that $H(y)=H(y')$ for any $y,y'\in Y$,
we shall denote by $H(Y)$ the energy of the states in $Y$.
For any $Y\subset X$ we shall denote by $F(Y)$ the set of the minima
of the energy inside $Y$, that is to say $y\in F(Y)$ if and only
if
$H(y')\ge H(y)$ for any $y'\in Y$.
We let $X_\rr{s}:=F(X)$ be the set of \definisco{ground states} of $H$,
namely, the set of the absolute minima of the energy.

For any positive integer $n$, $\omega\in X^n$
such that $(\omega_i,\omega_{i+1})\in Q$ for all $i=1,\dots,n-1$
is called a \definisco{path} joining $\omega_0$ to $\omega_n$;
we also say that $n$ is the length of the path.
For any path $\omega$ of length $n$, we let
\begin{equation}
\label{height}
\Phi_\omega:=\max_{i=1,\dots,n-1}[H(\omega_i)+\Delta(\omega_i,\omega_{i+1})]
\end{equation}
be the \definisco{height} of the path\footnote{Since the energy landscape
is reversible, the energy of the state $\omega_n$ is implicitly taken
into account in \eqref{height}, indeed \eqref{rev03} implies
$H(\omega_n)\le\Delta(\omega_{n-1},\omega_n)+H(\omega_{n-1})$.}.
For any $y,z\in X$ we denote by $\Omega(y,z)$
the set of the paths joining $y$ to $z$.
For any $y,z\in X$ we
define the \definisco{communication height}
between $y$ and $z$ as
\begin{equation}
\label{communication}
\Phi(y,z)
:=
\min_{\omega\in\Omega(y,z)}
\Phi_\omega
\end{equation}
From \eqref{rev03}, \eqref{height}, and \eqref{communication}
it follows immediately that
\begin{equation}
\label{rev02}
\Phi(y,z)=\Phi(z,y)
\end{equation}
for all $y,z\in X$.
For any $Y,Z\subset X$ we let
\begin{equation}
\label{communication-set}
\Phi(Y,Z)
:=
\min_{\omega\in\Omega(Y,Z)}\Phi_\omega
=
\min_{y\in Y,z\in Z}\Phi(y,z)
\end{equation}
where we have used the notation
$\Omega(Y,Z)$ for the set of paths joining
a state in $Y$ to a state in $Z$.

For any $y,z\in X$ we define also the
\definisco{communication cost} from
$y$ to $z$ as the quantity $\Phi(y,z)-H(y)$. Note that in general
the communication cost from $y$ to $z$ differs from that
from $z$ to $y$.

\subsection{Metastable states}
\label{s:meta}
\par\noindent
For any $x\in X$
we denote by $I_x$ the set of states $y\in X$
such that $H(y)<H(x)$.
Note that $I_x=\emptyset$ if $x\in X_\rr{s}$.
We then define the
\definisco{stability level} of any $x\in X\setminus X_\rr{s}$
\begin{equation}
\label{stability}
V_x:=\Phi(x,I_x)-H(x)
\ge0
\end{equation}
Note that the stability level $V_x$ of $x$ is the minimal
communication cost that, starting from $x$, has to be payed
in order to reach states at energy lower than $H(x)$.

Following \cite{MNOS} we now introduce the notion of
maximal stability level.
Assume $X\setminus X_\rr{s}\neq\emptyset$, we let
the \definisco{maximal stability level} be
\begin{equation}
\label{gamma}
\Gamma_\rr{m}:=\sup_{x\in X\setminus X_\rr{s}}V_x
\end{equation}
\begin{definition}
\label{def1}
 We call metastable set $X_{\rr{m}}$, the set
\begin{equation}
\label{metastabile}
X_\rr{m}
:=
\{x\in X\setminus X_\rr{s}:\,V_x=\Gamma_{\rr{m}}\}
\end{equation}
\end{definition}

\medskip

\noindent Note that, since the state space is finite, the maximal stability level
$\Gamma_\rr{m}$ is a finite number.
Following \cite{MNOS}, that is to say by assuming the so called
\emph{pathwise} point of view, we shall call
$X_\rr{m}$ the set of \definisco{metastable} states of the
system. Each state $x\in X_\rr{m}$ is called \definisco{metastable}.

A different, even if strictly related, notion of metastable states
is that given in
\cite{BEGK} in the framework of the Potential Theoretic Approach.
First recall that
the \textit{Dirichlet form} associated with the reversible Markov chain
is defined as the functional
\begin{equation}
\label{diri-new}
\mathscr{D}_\beta[f]
:=
\frac{1}{2}\sum_{y,z\in X}
 \mu_\beta(y)p_\beta(y,z) [f(y)-f(z)]^2
\end{equation}
where $f:X\to\bb{R}$ is a generic function.

Thus,
given two not empty disjoint sets $Y,Z\subset X$ the \emph{capacity}
of the pair $Y$ and $Z$ can be defined as
\begin{equation}
\label{capac}
\rr{cap}_\beta(Y,Z)
:=
\min_{\newatop{f:X\to[0,1]}{f\vert_Y=1,f\vert_Z=0}}
\mathscr{D}_\beta[f]
\end{equation}
Note that
the capacity is a \emph{symmetric} function of the sets $Y$ and $Z$.
It can be proven that the
right hand side of (\ref{capac}) has a unique minimizer
called \emph{equilibrium potential} of the pair $Y$ and $Z$ and
denoted by $h_{Y,Z}$.

A nice interpretation of the equilibrium potential
in terms of hitting times can be given.
For $x\in X$ and $Y\subset X$ we shall denote by
$\tau^x_Y$ the \definisco{first hitting} time to $Y$ of the chain started
at $x$. Whenever possible we shall drop the superscript denoting
the starting point from the notation.
Then, it can be proven that
\begin{equation}
\label{eqpot}
h_{Y,Z}(x)=
\left\{
\begin{array}{ll}
\bb{P}_x(\tau_Y<\tau_Z) & \;\;\textrm{ for } x\in X\setminus(Y\cup Z)\\
1& \;\;\textrm{ for } x\in Y\\
0& \;\;\textrm{ for } x\in Z\\
\end{array}
\right.
\end{equation}
where $\tau_Y$ and $\tau_Z$ are, respectively,
the first hitting time to $Y$ and $Z$ for the chain started at $x$.
It can be proven
that, for any $Y\subset X$ and $z\in X\setminus Y$,
\begin{equation}
\label{cap-prop}
\rr{cap}_\beta(z,Y)=\mu_\beta(z)\bb{P}_z(\tau_Y<\tau_z)
\end{equation}
see \cite[equation (3.10)]{Bo}.

\begin{definition}
\label{pta}
A set $M\subset X$ is said to be
{\em p.t.a.--metastable}
if
\begin{equation}
\label{metadef}
\lim_{\beta\to\infty}
\frac{\max_{x\notin{M}}\mu_\beta(x){[\rr{cap}_\beta(x,M)]}^{-1}}
     {\min_{x\in{M}}\mu_\beta(x){[\rr{cap}_\beta(x,M\setminus\{x\})]}^{-1}}
=0
\end{equation}
\end{definition}
The prefix p.t.a. stands for potential theoretic approach. We used
this expression in order to avoid confusion with the set of
metastable states $X_\rr{m}$ introduced in \eqref{metastabile}.
The physical meaning of the above definition
can be understood once one remarks that
the quantity $\mu_\beta(x)/\textrm{cap}_\beta(x,y)$,
for any $x,y\in X$, is strictly related to the communication
cost between the states $x$ and $y$, see Proposition~\ref{t:apriori}.
Thus, condition \eqref{metadef} ensures that the communication
cost between any state outside $M$ and $M$ itself is smaller
than the communication cost between any two states in $M$.
In other words, it states that
getting to $M$ starting from any state outside $M$ is
``much'' easier than going from any point in $M$ to any other
point in $M$.

Finally, given a p.t.a.--metastable set $M\subset X$,
for any $x\in M$ we let
\begin{equation}
\label{valley}
A(x):=\{y\in X:\bb{P}_y(\tau_x=\tau_M)=\sup_{z\in M}
\bb{P}_y(\tau_z=\tau_M)\}
\end{equation}
be the \definisco{valley} associated with $x$.
For any $y\in X$ and any $z\in M$ the quantity
$\bb{P}_y(\tau_z=\tau_M)$ measures the probability that, starting from
$y$, the system touches $M$ for the first time in $z$.
Thus, by computing
$\sup_{z,\in M}\bb{P}_y(\tau_z=\tau_M)$,
one detects the best way to touch $M$ for the system started at $y$.
Hence, the condition
$\bb{P}_y(\tau_x=\tau_M)=\sup_{z,\in M}\bb{P}_y(\tau_z=\tau_M)$
selects all the sites $y$ such that, for the system started
at $y$, the best way to touch $M$ is that of touching it for
the first time at $x$.

\subsection{Series of metastable states}
\label{s:series}
\par
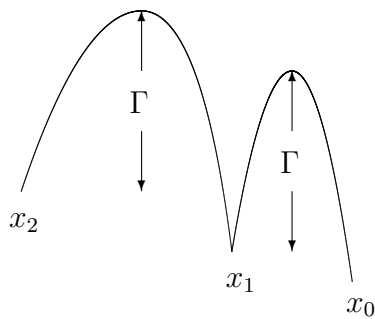
\begin{figure}[t]
 \begin{picture}(100,130)(-30,25)
 \thinlines
 \setlength{\unitlength}{0.08cm}
 \qbezier(50,30)(60,60)(70,60)
 \qbezier(70,60)(80,60)(85,20)
 \put(70,50){\vector(0,1){10}}
 \put(70,40){\vector(0,-1){10}}
 \put(68,43){${\Gamma}$}
 \put(48,24){${\metadue}$}
 \put(84,14){${\metauno}$}
 \qbezier(85,20)(90,50)(95,50)
 \qbezier(95,50)(100,50)(105,15)
 \put(95,40){\vector(0,1){10}}
 \put(95,30){\vector(0,-1){10}}
 \put(93,33){${\Gamma}$}
 \put(104,10){${\stab}$}
 \end{picture}
 \caption{Schematic description of the energy landscape for a series of
          metastable states.}
 \label{fig:fig01}
\end{figure}

\noindent
The aim of this paper is that of proving an addition formula for
the exit time from metastable states in the case in which they
form a series. With this expression we mean that the structure of the energy
landscape is such that the system has two non degenerate in
energy metastable states and the system, started at the one
having higher energy,
must necessarily pass through the second one before relaxing to the
stable state. See Fig.~\ref{fig:fig01} for a schematic description
of the situation we have in mind and that will be
formalized through the following
conditions.

\begin{condition}
\label{t:series00}
Recall (\ref{gamma}) and (\ref{metastabile}).We assume that the energy landscape $(X,Q,H,\Delta)$ is such that
there exist three states $\metadue$, $\metauno$, and $\stab$
such that
$X_\rr{s}=\{\stab\}$,
$X_\rr{m}=\{\metauno,\metadue\}$,
and
$H(\metadue)>H(\metauno)$.
\end{condition}

Note that, by recalling the definition of the set of ground states $X_\rr{s}$,
we immediately have that
\begin{equation}
\label{msl0-1}
H(x_1)>H(x_0)
\end{equation}
Moreover, from the definition \eqref{gamma}
of maximal stability level
it follows that (see \cite[Theorem~2.3]{CN2013})
the communication cost from $\metadue$ to $\stab$ is equal
to that from $\metauno$ to $\stab$, that is to say
\begin{equation}
\label{msl00}
\Phi(x_2,x_0)-H(x_2)
=
\Phi(x_1,x_0)-H(x_1)
=
\Gamma_\rr{m}
\end{equation}
Note that, since
$x_2$ is a metastable state, its stability level
cannot be lower than $\Gamma_\rr{m}$.
Then, recalling that $H(x_2)>H(x_1)$, one has that
$\Phi(x_2,x_1)-H(x_2)\ge\Gamma_\rr{m}$. On the other hand,
\eqref{msl00} implies that there exists a path $\omega\in\Omega(x_2,x_1)$
such that $\Phi_\omega=H(x_2)+\Gamma_\rr{m}$ and, hence,
$\Phi(x_2,x_1)-H(x_2)\le\Gamma_\rr{m}$. The two bounds finally imply that
\begin{equation}
\label{series01new}
\Phi(x_2,x_1)-H(x_2)
=
\Gamma_\rr{m}
\end{equation}
Note that
the communication cost from
$\stab$ to $\metadue$ and that from
$\metauno$ to $\metadue$
are larger than $\Gamma_\rr{m}$, that is to say,
\begin{equation}
\label{indietro}
\Phi(\stab,\metadue)-H(\stab)
>
\Gamma_\rr{m}
\;\;\;\textrm{ and }\;\;\;
\Phi(\metauno,\metadue)-H(\metauno)
>
\Gamma_\rr{m}
\end{equation}
Indeed, by recalling the
reversibility property \eqref{rev02}
we have
\begin{eqnarray*}
\Phi(\metauno,\metadue)-H(\metauno)
&=&
\Phi(\metadue,\metauno)
-H(\metadue)
+H(\metadue)-H(\metauno)\\
&=&
\Gamma_\rr{m}
+H(\metadue)-H(\metauno)
>
\Gamma_\rr{m}
\end{eqnarray*}
where in the last two steps we have used
\eqref{series01new} and Condition~\ref{t:series00}, which proves the second
of the two equations \eqref{indietro}. The first of them can
be proved similarly.

We want to implement in the
model the series structure depicted in Fig.~\ref{fig:fig01}.
With this we mean that when the system is started at $x_2$ with high
probability it will visit $x_1$ before $x_0$. For this reason we shall
assume the following condition.

\begin{condition}
\label{t:series01}
Condition~\ref{t:series00} is satisfied and
\begin{equation}
\label{series01}
\lim_{\beta\to\infty}\bb{P}_{\metadue}(\tau_\stab<\tau_\metauno)=0
\end{equation}
\end{condition}
We remark that the Condition~\ref{t:series01} is indeed a condition
on the equilibrium potential $h_{x_0,x_1}$ evaluated at $x_2$.

The most important goal of this paper is that of proving the
formula \eqref{addition02} for the expectation of the escape time
$\tau_\stab$ for the chain started at $\metadue$.
Such an expectation, hence, will be of order $\exp\{\beta\Gamma_\rr{m}\}$
and the prefactor will be that given in \eqref{addition02}.
At the level of logarithmic equivalence, namely, by renouncing
to get sharp estimate, this result can be proven by
the methods in \cite{MNOS}. More precisely, one gets that
$(1/\beta)\log\bb{E}_\metadue[\tau_\stab]$ tends to $\Gamma_\rr{m}$
in the large $\beta$ limit.


We can thus formulate the further assumptions that we shall need
in the sequel in order to discuss the problem from the point of
view of the Potential Theoretic Approach.

\begin{condition}
\label{t:series02}
Condition~\ref{t:series00} is satisfied and
there exists two positive constants $k_1,k_2<\infty$ such that
\begin{equation}
\label{series02}
\frac{\mu_\beta(\metadue)}{\rr{cap}_\beta(\metadue,\{\metauno,\stab\})}
=
\frac{1}{k_2}
e^{\beta\Gamma_\rr{m}}[1+o(1)],\,\,\,\,\,
\frac{\mu_\beta(x_1)}{\rr{cap}_\beta(\metauno,\stab)}
=
\frac{1}{k_1}
e^{\beta\Gamma_\rr{m}}[1+o(1)]
\end{equation}
where $o(1)$ denotes a function tending to zero in the limit
$\beta\to\infty$.
\end{condition}

\subsection{Main results}
\label{s:mainris}
\par\noindent
We shall prove the addition rule for the exit times from the metastable
states by using the sharp estimates provided by the Potential
Theoretic Approach to metastability originally
developed in \cite{BEGK}.

\begin{theorem}
\label{t:ptaset}
Assume Conditions~\ref{t:series00} is satisfied.
Then $\{x_0,x_1,x_2\}\subset X$ is a p.t.a.--metastable set.
\end{theorem}

By means of the theory in \cite{BEGK}
it is possible to write asymptotic estimates of
the first hitting time to a subset of a p.t.a.--metastable set
when the dynamics is started in state of the same p.t.a.--metastable
set not belonging to the considered subset. These results, see for instance
\cite[Theorem 1.3]{BEGK}, are typically proven
under suitable not degeneracy conditions \cite[Definition 1.2]{BEGK} that
are not satisfied in our
case, due to the presence of multiple metastable states.
In the following theorem
we state two results that, for the reasons outlined above,
can be deduced directly from those in \cite{BEGK}.
On the other hand, as we shall discuss in detail in
Section~\ref{s:dimo}, they can be deduced by
some of the results proven in
\cite{Bo} (see, also, \cite{BdH}).
But,
since we assumed strong hypotheses on the energy landscape of the
model, it will be possible to prove the theorem
directly by
means of simple estimates.
This ``ad hoc'' proof is also given in
Section~\ref{s:dimo}.

\begin{theorem}
\label{t:meant}
Assume Conditions~\ref{t:series00} is satisfied. Then
\begin{equation}
\label{meant}
\mathbb{E}_{\metadue}[\tau_{\{\metauno,\stab\}}]\!=\!
\frac{\mu_\beta(\metadue)}
     {\rr{cap}_\beta(\metadue,\{\metauno,\stab\})}[1+o(1)],\,
\mathbb{E}_{\metauno}[\tau_\stab]\!=\!
\frac{\mu_\beta(\metauno)}{\rr{cap}_\beta(\metauno,\stab)}[1+o(1)]
\end{equation}
\end{theorem}

\begin{theorem}
\label{t:addition01}
Assume Conditions~\ref{t:series00}
and \ref{t:series02} are satisfied.
Then
\begin{equation}
\label{addition01}
\bb{E}_\metadue[\tau_{\{\metauno,\stab\}}]
=
e^{\beta\Gamma_\rr{m}}\frac{1}{k_2}[1+o(1)]
\;\;\;\textrm{ and }\;\;\;
\bb{E}_\metauno[\tau_\stab]
=
e^{\beta\Gamma_\rr{m}}\frac{1}{k_1}[1+o(1)]
\end{equation}
\end{theorem}

\begin{theorem}
\label{t:addition02}
Assume Conditions~\ref{t:series00}, \ref{t:series01},
 and \ref{t:series02} are satisfied.
Then
\begin{equation}
\label{addition02}
\bb{E}_\metadue[\tau_\stab]
=
e^{\beta\Gamma_\rr{m}}\Big(\frac{1}{k_1}+\frac{1}{k_2}\Big)[1+o(1)]
\end{equation}
\end{theorem}

We remark that Theorem~\ref{t:addition02} gives an addition formula
   for the mean first hitting time to $\stab$ starting from $\metadue$.
   Neglecting terms of order $o(1)$, such a mean time can be written as
   the sum of the mean hitting time to the pair $\{\metauno,\stab\}$ when
   the chain is started at $\metadue$ and of the mean hitting time to
   $\stab$ when the chain is started at $\metauno$.
   It is very interesting to note that no role is plaid in this
   decomposition by the mean hitting time to $\metauno$ for the chain
   started at $\metadue$. Indeed, on the $\exp\{\beta\Gamma_\rr{m}\}$ time
   scale no control can be proven for such a
   mean time. For instance, in the case of the Blume--Capel model that
   will be studied in Section~\ref{s:blume}, in
   \cite[Proposition~2.5]{LL2016} it is proven that
   $\mathbb{E}_{\metadue}[\tau_\metauno]/e^{\beta \Gamma_\textnormal{m}}$
   diverges in the limit $\beta\to\infty$.

\section{Proof of results}
\label{s:dimo}
\par\noindent
In this section we proof the theorems stated above and
related to the general setup given in Section~\ref{s:series}.

\medskip
\par\noindent
\textit{Proof of Theorem~\ref{t:ptaset}.\/}
The theorem follows immediately by Condition~\ref{t:series00},
\eqref{series01new}, and \cite[Theorem~3.6]{CN2013}.
\qed

We just note that the Theorem~3.6 in \cite{CN2013} has been
proved in a slightly different context,
but the proof given there
applies also to the more general case studied here.

Theorem~\ref{t:meant} can be deduced by using the structure provided
by Theorem~\ref{t:ptaset} above, and the general results
\cite[Eq. (4.14) and Lemma 4.3]{Bo}. Alternatively, one can use
\cite[Eq. (8.1.6), Eq. (8.3.3), and Lemma 8.13]{BdH}.
Since we assumed strong hypotheses on the energy landscape of the
model, it is possible to prove directly equation \eqref{meant} by
means of simple estimates.
Before discussing such a proof we
state two useful lemmas.
Recall Condition~\ref{t:series00},
in the first of the two lemmas we collect two bounds to the
energy cost that has to be payed to go from any state $x\neq \metauno$
to $\metauno$ or to $\stab$.
The second lemma is similar.

\begin{lemma}
\label{stimePhi}
Assume Condition~\ref{t:series00} is satisfied. For any $x\in X$ and
$x\neq\metauno$.
If $H(x)\leq H(\metauno)$, we have that
\begin{equation}
\label{claim1}
\Phi(x,\stab)-H(x)
<\Gamma_\rr{m}
\;\;\textrm{ and }\;\;
\Phi(x,\metauno)
-H(\metauno)
\ge \Gamma_\rr{m}
\end{equation}
\end{lemma}
\medskip
\par\noindent
\textit{Proof.\/}
Let us prove the first inequality. 
By Theorem~2.3  in \cite{CN2013} we have that
$\Phi(x,\stab)\leq\Gamma_\rr{m}+H(x)$.
If by contradiction $\Phi(x,\stab)=\Gamma_\rr{m}+H(x)$ then, by the same
Theorem~2.3 in \cite{CN2013}, $x\in X_\rr{m}$ which is in contradiction
with Condition~\ref{t:series00}.

As regards the proof of the second inequality we distinguish two cases.
Case $H(x)<H(\metauno)$:
we have that $x\in I_\metauno$.
By Definition~\ref{def1} of metastable state and by (\ref{stability}), we get
$$\Phi(\metauno,x)\geq \Phi(\metauno,I_{\metauno})= \Gamma_\rr{m}+H(\metauno)$$
that proves the inequality.

Case $H(x)=H(\metauno)$:
let us define the set
\begin{displaymath}
\mathcal{C}:=\{y\in X: \Phi(y,\metauno)<H(\metauno)+\Gamma_\rr{m}\}
\end{displaymath}
and show that $x\not\in \mathcal{C}$. Since $H(x)= H(\metauno)$,  the identity $I_{x}=I_\metauno$ follows. Furthermore, being $\metauno\in X_\rr{m}$, we have $\mathcal{C}\cap I_{\metauno}=\emptyset$; hence,
 $\mathcal{C}\cap I_{x}=\emptyset$ as well.
Moreover,  if $x\in\mathcal{C}$ then $V_x=\Phi(x,I_x)-H(x)\geq H(\metauno)+\Gamma_\rr{m}-H(x)=\Gamma_\rr{m}$. By the Definition~\ref{def1},  $x$ would be a metastable state, in contradiction with Condition~\ref{t:series00}.
Hence, since $x\not\in \mathcal{C}$, we have that
$$
\Phi(x,\metauno)\geq \Gamma_\rr{m}+H(\metauno)
$$
that proves the inequality. the inequality.
\qed
\begin{lemma}
\label{stimePhi2}
Assume Condition~\ref{t:series00} is satisfied. For any $x\in X$ and
$x\notin\{\metadue,\metauno,\stab\}$.
If $H(x)\leq H(\metadue)$, we have that
\begin{equation}
\label{claim3}
\Phi(x,\{\metauno,\stab\})-H(x)
<\Gamma_\rr{m}
\;\;\textrm{ and }\;\;
\Phi(x,\metadue)
-H(\metadue)
\ge \Gamma_\rr{m}
\end{equation}
\end{lemma}
\medskip
\par\noindent
\textit{Proof.\/}
Let us prove the first inequality.
By Theorem~2.3  in \cite{CN2013} we have that
$\Phi(x,\{x_1,\stab\})\leq \Phi(x,\stab)\leq\Gamma_\rr{m}+H(x)$.
If by absurdity $\Phi(x,\stab)=\Gamma_\rr{m}+H(x)$ then, by the same
Theorem~2.3 in \cite{CN2013}, $x\in X_\rr{m}$ which is in contradiction
with Condition~\ref{t:series00}.

As regards the proof of the second inequality we distinguish two cases.
Case $H(x)<H(\metadue)$:
we have that $x\in I_\metadue$.
By Definition~\ref{def1} of metastable state and by (\ref{stability}), we get
$$\Phi(\metadue,x)\geq \Phi(\metadue,I_{\metadue})= \Gamma_\rr{m}+H(\metadue)$$
that proves the inequality.

Case $H(x)=H(\metadue)$:
let us define the set
\begin{displaymath}
\mathcal{C}:=\{y\in X: \Phi(y,\metadue)<H(\metadue)+\Gamma_\rr{m}\}
\end{displaymath}
and show that $x\not\in \mathcal{C}$. Since $H(x)= H(\metadue)$,  the identity $I_{x}=I_\metadue$ follows. Furthermore, being $\metadue\in X_\rr{m}$, we have $\mathcal{C}\cap I_{\metadue}=\emptyset$; hence,
 $\mathcal{C}\cap I_{x}=\emptyset$ as well.
Moreover,  if $x\in\mathcal{C}$ then $V_x=\Phi(x,I_x)-H(x)\geq H(\metadue)+\Gamma_\rr{m}-H(x)=\Gamma$. By the Definition~\ref{def1},  $x$ would be a metastable state, in contradiction with Condition~\ref{t:series00}.
Hence, since $x\not\in \mathcal{C}$, we have that
$$
\Phi(x,\metadue)\geq \Gamma_\rr{m}+H(\metadue)
$$
that proves the inequality.
\qed

\medskip
\par\noindent
\textit{Proof of Theorem~\ref{t:meant}.\/}
 We prove in details the right of equation (\ref{meant}).
 The proof is based on Lemma~\ref{stimePhi}. The equation on the left can be deduced with precisely the same arguments and using the bounds in Lemma~\ref{stimePhi2}. The only general results used is the representation of the expected mean time in terms of the Green function given in \cite[Corollary~3.3]{BEGK} (see also equation (3.18) in
the proof of the Theorem~3.5 in \cite{BEGK} or \cite[Eq. (4.29)]{Alex}).
Indeed, recalling \eqref{cap-prop} above,
we have:
\begin{equation}
\label{meant1}
\mathbb{E}_{\metauno}[\tau_\stab]
= \frac{1}{\rr{cap}_\beta(\metauno,\stab)}
\sum_{x\in X}
\mu_\beta(x) \, h_{\metauno,\stab}(x)
\end{equation}
Considering the contribution of $\metauno$ in the sum and
recalling \eqref{eqpot}, we get the following lower bound:
\begin{equation}
\label{meant_lower}
\mathbb{E}_{\metauno}[\tau_\stab]
\ge \frac{1}{\rr{cap}(\metauno,\stab)} \mu_\beta(\metauno) h_{\metauno,\stab}(\metauno)=
\frac{1}{\rr{cap}(\metauno,\stab)} \mu_\beta(\metauno)
\end{equation}
In order to provide un upper bound, we first use the
boundary conditions in \eqref{eqpot} to rewrite \eqref{meant1} as follows:
\begin{displaymath}
\mathbb{E}_{\metauno}[\tau_\stab]
=
\frac{1}{\rr{cap}(\metauno,\stab)}
\Big[
\sum_{{x\in X\setminus \stab,}\atop{H(x)\leq H(\metauno)}}
\mu_\beta(x) h_{\metauno,\stab}(x)+
\sum_{{x\in X\setminus \stab,}\atop{H(x)> H(\metauno)}}
\mu_\beta(x) h_{\metauno,\stab}(x)
\Big]
\end{displaymath}
Recalling that
$h_{\metauno,\stab}(\metauno)=1$, the equilibrium potential is not bigger
than one, the configuration space is finite, and
$\mu_\beta(x)=\mu_\beta(\metauno)\exp\{-\beta \delta\}$ for some positive
$\delta$ and for any $x\in X$ such that $H(x)>H(\metauno)$, we get
\begin{equation}
\label{inpiu}
\mathbb{E}_{\metauno}[\tau_\stab]
=
\frac{1}{\rr{cap}(\metauno,\stab)}
\Big[
\!\!
\!\!
\sum_{{x\in X\setminus \stab,}\atop{H(x)\leq H(\metauno),\,x\neq\metauno}}
\!\!
\!\!
\mu_\beta(x)\,h_{\metauno,\stab}(x)+
\mu_\beta(\metauno)[1+o(1)]
\Big]
\end{equation}

By (\ref{eqpot}) and (\ref{potbound}) we can give the following upper bound
for the equilibrium potential $h_{\metauno,\stab}(x)$, for any
$x\neq\metauno,\stab$
\begin{displaymath}
h_{\metauno,\stab}(x)\leq\frac{\rr{cap}(x,\metauno)}{\rr{cap}(x,\stab)}
\;\;.
\end{displaymath}
Thus, if $H(x)\le H(\metauno)$, we have
\begin{eqnarray*}
h_{\metauno,\stab}(x)
&\leq&
C
\frac{e^{-\beta\Phi(x,\metauno)}}{e^{-\beta\Phi(x,\stab)}}\leq C
\frac{e^{-\beta(\Gamma_\rr{m}+H(\metauno))}}{e^{-\beta(\Gamma_\rr{m}+H(x)-\delta)}}
=
 C e^{-\beta\delta}\frac{\mu_\beta(\metauno)}{\mu_\beta(x)}
\end{eqnarray*}
where in the first inequality we used Proposition~\ref{t:apriori},
in the second Lemma~\ref{stimePhi}, and
$C,\delta$ are suitable positive constants.
By using \eqref{inpiu} we get
\begin{displaymath}
\mathbb{E}_{\metauno}[\tau_\stab]
\le
\frac{1}{\rr{cap}(\metauno,\stab)}
\Big[
\!\!
\!\!
\sum_{{x\in X\setminus \stab,}\atop{H(x)\leq H(\metauno),\,x\neq\metauno}}
\!\!
\!\!
C\mu_\beta(x) e^{-\beta\delta}\frac{\mu_\beta(\metauno)}{\mu_\beta(x)}+
\mu_\beta(\metauno)[1+o(1)]
\Big]
\end{displaymath}
Which implies
\begin{equation}
\label{uppot}
\mathbb{E}_{\metauno}[\tau_\stab]
\le
\frac{\mu_\beta(\metauno)}{\rr{cap}(\metauno,\stab)}[1+o(1)]
\end{equation}
where we have used that the configuration space is finite.
The Theorem finally follows by \eqref{meant_lower} and \eqref{uppot}.
\qed

\medskip
\par\noindent
\textit{Proof of Theorem~\ref{t:addition01}.\/}
The theorem follows immediately by exploiting
Condition~\ref{t:series02} and applying
Theorem~\ref{t:meant}.
\qed

The proof of Theorem~\ref{t:addition02} is based on the following lemma.

\begin{lemma}
\label{t:dimo00}
Given three states $y,w,z\in X$ pairwise mutually different, we have
that the following holds
\begin{equation}
\label{dimo00}
\bb{E}_y[\tau_z]
=
\bb{E}_y[\tau_{\{w,z\}}]
+
\bb{E}_w[\tau_z]\bb{P}_y(\tau_w<\tau_z)
\end{equation}
\end{lemma}

\medskip
\par\noindent
\textit{Proof.\/}
First of all we note that
\begin{eqnarray*}
\bb{E}_y(\tau_z)&=&\bb{E}_y[\tau_z\bb{I}_{\tau_w<\tau_z}]+\bb{E}_y[\tau_z\bb{I}_{\tau_w\geq\tau_z}]
\end{eqnarray*}

%
%
\par\noindent
We now rewrite the first term as follows

\begin{eqnarray*}
\bb{E}_y[\tau_z\bb{I}_{\{\tau_w<\tau_z\}}]
&=&\bb{E}_y[ \bb{E}_y[\tau_z\bb{I}_{\{\tau_w<\tau_z\}}|\mathcal{F}_{\tau_w}]
]
=\bb{E}_y[\bb{I}_{\{\tau_w<\tau_z\}}(\tau_w+\bb{E}_w[\tau_z])]\\
&=& \bb{E}_y[\tau_w\bb{I}_{
\{\tau_w<\tau_z\}}]+\bb{P}_y(\tau_w<\tau_z)\bb{E}_w(\tau_z).
\end{eqnarray*}
where we have used the fact that $\tau_w$ is a stopping time, that $\bb{I}_{\{\tau_w<\tau_z\}}$ is measurable with respect to the pre--$\tau_w$--$\sigma$--algebra $\mathcal{F}_{\tau_w}$ 
and the strong Markov property which gives $\bb{E}_y[\tau_z|\mathcal{F}_{\tau_w}]=\tau_w+
\bb{E}_w[\tau_z]$ on the event $\{\tau_w\leq \tau_z\}$.
Since $ (\tau_w\mathbb{I}_{\{ \tau_w<\tau_z \}}+\tau_z\mathbb{I}_{\{ \tau_w\geq\tau_z \}})=\tau_{\{w,z\}}$,
(\ref{dimo00})  follows.
\qed

\medskip
\par\noindent
\textit{Proof of Theorem~\ref{t:addition02}.\/}
By \eqref{dimo00} we have that
\begin{displaymath}
\bb{E}_\metadue[\tau_\stab]
=
\bb{E}_\metadue[\tau_{\{\metauno,\stab\}}]
+
\bb{E}_\metauno[\tau_\stab]\bb{P}_\metadue(\tau_\metauno<\tau_\stab)
\end{displaymath}
By Theorem~\ref{t:addition01} and Condition~\ref{t:series01} it follows that
$$
{\bb{E}_\metadue[\tau_\stab]}=
     {e^{\beta\Gamma_\rr{m}}\left(\frac{1}{k_1}+\frac{1}{k_2}\right)}
[1+o(1)]
$$
\qed


\section{Application to the Blume--Capel model}
\label{s:blume}
\par\noindent
In this section, as a possible application of the theory described
above, we apply our results to the case of
the Blume--Capel model (\cite{B,C}). In particular,  we consider the model   with null chemical potential,
which has
two metastable states non degenerate in energy.
We shall then derive, in a different way, the results already
appeared in \cite{LL2016}.

Let us consider a square lattice $\Lambda\subset \mathbb{Z}^2$
with periodic boundary conditions and side length $L$.
Let $\{-1,0,+1\}$
be the single spin state space and $\mathcal{X}:=\{-1,0,+1\}^{\Lambda}$
be the configuration space. The Hamiltonian of
the model \cite{CN2013} is
\begin{equation}
\label{cap:002}
H(\sigma)=\sum_{\langle i,j\rangle}(\sigma(i)-\sigma(j))^2
-h \sum_{i\in\Lambda}\sigma(i)
\end{equation}
for any $\sigma\in\mathcal{X}$, where the first sum runs over
the pairs of nearest neighbors
and $h\in\mathbb{R}$ is the magnetic field.
We denote by $\mu_\beta$ the corresponding Gibbs measure
\begin{displaymath}
\mu_\beta(\sigma)=\exp\{-\beta H(\sigma)\}
  /\sum_{\eta\in\cc{X}} \exp\{-\beta H(\eta)\}
\end{displaymath}
with inverse temperature $\beta$.
We shall study
the zero chemical potential
Blume--Capel model for the following choice of the parameters:
the magnetic field $h$ and the torus $\Lambda$ are such that
$0<h<1$, $2/h$ is not integer, and $|\Lambda|\ge49/h^4$
finite, where,
for any positive real $a$, we let $\lfloor a\rfloor$ be the largest
integer smaller than or equal to $a$.

The time evolution of the model is defined by the Metropolis
Markov chain
$\sigma_t$ with $t=0,1,\ldots$ the discrete time variable,
see Section~\ref{s:esempi}, with Hamiltonian $H$ and
connectivity matrix
$$
q(\sigma,\eta):=\left\{
\begin{array}{ll}
0& \textnormal{if } \sigma, \eta \textnormal{ differ at more than one site}
\\
1/(2|\Lambda |)
& \textnormal{otherwise } 
\end{array}
\right.
.
$$

As already remarked in Section~\ref{s:esempi}, the dynamics above
is an example of the dynamics defined in Section~\ref{s:reversibile},
provided we let
$i)$ $r(x,x)=0$ and
$r(x,y)=-\log q(x,y)$ for $x\neq y$ and $q(x,y)>0$;
$ii)$
for any $x\neq y$,
$\Delta(x,y)=\infty$ if $q(x,y)=0$ and
$\Delta(x,y)=[H(y)-H(x)]_+$ otherwise;
$iii)$
$\Delta(x,x)=\infty$ if $p_\beta(x,x)=0$ and
$\Delta(x,x)=-(1/\beta)\log p_\beta(x,x)$ otherwise.
The notation introduced in Section~\ref{s:reversibile}--\ref{s:meta}
is then trivially particularized to the Blume--Capel case.

Given $V\subset\Lambda$ and $\sigma\in\cc{X}$, we let $\sigma_V$ be
the restriction of $\sigma$ to $V$, namely,
$\sigma_V\in\{-1,+1\}^V$ such that $\sigma_V(i)=\sigma(i)$
for any $i\in V$.

We let $\puno\in \cc{X}$ to be the configuration such that
$\puno(i)=+1$ for all $i\in\Lambda$.
Other two very relevant configurations are $\muno$ and
$\zero$, that is the configuration in which all the spin are
minus one and the one in which all the spins are zero.
Note that in these configurations the exchange part of the energy
is minimal, although the magnetic part is not.

\begin{figure}[t]\scalebox{.80}{
 \begin{picture}(100,90)(0,15)
 \thinlines
 \setlength{\unitlength}{0.045cm}
 \put(60,15){\line(1,0){40}}
 \put(60,15){\line(0,1){45}}
 \put(60,60){\line(1,0){40}}
 \put(100,15){\line(0,1){10}}
 \put(100,60){\line(0,-1){30}}
 \put(100,25){\line(1,0){5}}
 \put(100,30){\line(1,0){5}}
 \put(105,25){\line(0,1){5}}
 \put(75,63){${\scriptstyle\ell_\rr{c}-1}$}
 \put(53,35){${\scriptstyle \ell_\rr{c}}$}
 \put(90,18){${\scriptstyle \pcri}$}
 \put(70,30){\begin{turn}{30}{{\scriptsize zeros}}\end{turn}}
 \put(105,50){\begin{turn}{-30}{\scriptsize minuses}\end{turn}}
 \put(150,15){\line(1,0){40}}
 \put(150,15){\line(0,1){45}}
 \put(150,60){\line(1,0){40}}
 \put(190,15){\line(0,1){10}}
 \put(190,60){\line(0,-1){30}}
 \put(190,25){\line(1,0){5}}
 \put(190,30){\line(1,0){5}}
 \put(195,25){\line(0,1){5}}
 \put(165,63){${\scriptstyle \ell_\rr{c}-1}$}
 \put(143,35){${\scriptstyle \ell_\rr{c}}$}
 \put(180,18){${\scriptstyle \qcri}$}
 \put(160,30){\begin{turn}{30}{\scriptsize pluses}\end{turn}}
 \put(195,50){\begin{turn}{-30}{\scriptsize zeros}\end{turn}}
 \put(240,15){\line(1,0){10}}
 \put(250,15){\line(0,-1){5}}
 \put(250,10){\line(1,0){5}}
 \put(255,10){\line(0,1){5}}
 \put(255,15){\line(1,0){25}}
 \put(240,15){\line(0,1){45}}
 \put(240,60){\line(1,0){40}}
 \put(280,15){\line(0,1){45}}
 \put(280,60){\line(0,-1){30}}
 \put(251,63){${\scriptstyle \ell_{\textrm{c}}-1}$}
 \put(231,35){${\scriptstyle \ell_{\textrm{c}}}$}
 \put(269,18){${\scriptstyle \pcri'}$}
 \put(250,30){\begin{turn}{30}{\scriptsize zeroes}\end{turn}}
 \put(285,50){\begin{turn}{-30}{\scriptsize minuses}\end{turn}}
 \put(330,15){\line(1,0){10}}
 \put(340,15){\line(0,-1){5}}
 \put(340,10){\line(1,0){5}}
 \put(345,10){\line(0,1){5}}
 \put(345,15){\line(1,0){25}}
 \put(330,15){\line(0,1){45}}
 \put(330,60){\line(1,0){40}}
 \put(370,15){\line(0,1){45}}
 \put(370,60){\line(0,-1){30}}
 \put(341,63){${\scriptstyle \ell_{\textrm{c}}-1}$}
 \put(321,35){${\scriptstyle \ell_{\textrm{c}}}$}
 \put(358,18){${\scriptstyle \qcri'}$}
 \put(340,30){\begin{turn}{30}{\scriptsize pluses}\end{turn}}
 \put(375,50){\begin{turn}{-30}{\scriptsize zeros}\end{turn}}
 \end{picture}}
 \caption{Schematic representation of the configurations
          $\pcri$ and $\qcri$.
          The protuberance can be either on the left
          or on the right vertical edge (the longest ones in the
          picture) and there it
          can be shifted freely.
          Note that the two configurations $\pcri'$ and $\qcri'$, with the
          protuberance on the shortest side, do not play any role in the
          transition from the metastable to the stable state.}
 \label{fig:fig02}
\end{figure}
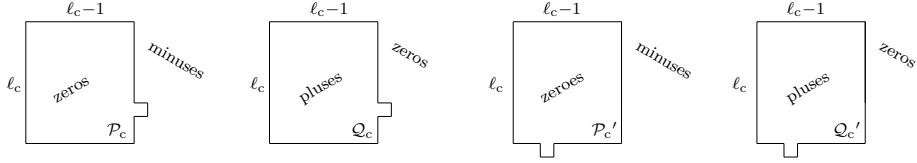

We now define the \definisco{critical length} of the model as
\begin{equation}
\label{lc}
\ell_\rr{c}:=\Big\lfloor\frac{2}{h}\Big\rfloor+1
\end{equation}
We denote by
$\pcri$ the set of configurations
in which all the spins are minus excepted those, which are zeros, in a
rectangle of sides long $\ell_\rr{c}$ and $\ell_\rr{c}-1$ and
in a site adjacent to one of the longest sides of the rectangle
(see Fig.~\ref{fig:fig02}).
We denote by
$\qcri$ the set of configurations
in which all the spins are zeros excepted those, which are pluses, in a
rectangle of sides long $\ell_\rr{c}$ and $\ell_\rr{c}-1$ and
in a site adjacent to one of the longest sides of the rectangle
(see the caption of Fig.~\ref{fig:fig02}).
We have:
\begin{displaymath}
H(\pcri)-H(\muno)
\!
=
\!
H(\qcri)-H(\zero)
\!
=
\!
4\ell_\rr{c}-h[\ell_\rr{c}(\ell_\rr{c}-1)+1]
\end{displaymath}
We then set
\begin{displaymath}
\Gamma_\rr{c}
:=
H(\pcri)-H(\muno)
=
H(\qcri)-H(\zero)
\end{displaymath}
A simple direct computation shows that for $h$ small one has
$\Gamma_\rr{c}\sim4/h$.

In order to give a result in the spirit of Theorem~\ref{t:addition02}
in the case of the Blume--Capel model, we first have to prove
preliminary Lemmas ensuring that the Conditions assumed in
the general discussion in Section~\ref{s:series} are satisfied
in the Blume--Capel case.

\begin{lemma}
\label{t:bc010}
With the parameters chosen as below \eqref{cap:002}, we have that
$\cc{X}_\rr{s}=\{\puno\}$,
$\cc{X}_\rr{m}=\{\muno,\zero\}$,
and
$H(\muno)>H(\zero)$.
Moreover, the maximal stability level $\Gamma_{\normalfont\textrm{m}}$
is equal to $\Gamma_{\normalfont\textrm{c}}$.
\end{lemma}

\begin{lemma}
\label{t:bc020}
With the parameters chosen as below \eqref{cap:002}, we have that
there exists $\kappa>0$ and $\beta_0>0$ such that for any $\beta>\beta_0$
\begin{equation}
\label{bc020}
\bb{P}_{\muno}(\tau_\puno<\tau_\zero)\le e^{-\beta\kappa}
\end{equation}
\end{lemma}


\begin{lemma}
\label{t:bc040}
With the parameters chosen as below \eqref{cap:002}, we have that
\begin{equation}
\label{bc040}
\frac{\mu_\beta(\muno)}{\rr{cap}_\beta(\muno,\{\zero,\puno\})}=
\frac{3e^{\beta\Gamma_{\normalfont\textrm{c}}}}
     {2(2\ell_{\normalfont\textrm{c}}-1)}\,
[1+o(1)],\,\,\,
\frac{\mu_\beta(\zero)}{\rr{cap}_\beta(\zero,\puno)}=
\frac{3e^{\beta\Gamma_{\normalfont\textrm{c}}}}
     {2(2\ell_{\normalfont\textrm{c}}-1)}\,
[1+o(1)]
\end{equation}
\end{lemma}

The Lemmas~\ref{t:bc010} and \ref{t:bc020} will be proven in
Section~\ref{s:proof-esempio} below.
The proof of the Lemma~\ref{t:bc040} will be given for completeness
in Section~\ref{s:proof-esempio}; but we stress that a completely
analogous result has been already proven in
\cite[Proposition~2.3]{LL2016} for the continuous time version
of the same model.

We finally state our main results about the sharp estimate
in the exit time in the Blume--Capel model with zero chemical
potential.

\begin{theorem}
\label{t:bc050}
With the parameters chosen as below \eqref{cap:002}, we have that
\begin{equation}
\label{bc050}
\bb{E}_\muno[\tau_{\{\zero,\puno\}}]
=
e^{\beta\Gamma_{\normalfont\textrm{c}}}
\frac{3}{2(2\ell_{\normalfont\textrm{c}}-1)}[1+o(1)]
,\,\,\,
\bb{E}_\zero[\tau_\puno]
=
e^{\beta\Gamma_{\normalfont\textrm{c}}}
\frac{3}{2(2\ell_{\normalfont\textrm{c}}-1)}[1+o(1)]
\end{equation}
\end{theorem}
\begin{theorem}
\label{t:bc060}
With the parameters chosen as below \eqref{cap:002}, we have that
\begin{equation}
\label{bc060}
\bb{E}_\muno[\tau_\puno]
=
e^{\beta\Gamma_{\normalfont\textrm{c}}}
\frac{3}{(2\ell_{\normalfont\textrm{c}}-1)}[1+o(1)]
\end{equation}
\end{theorem}

The proof of the theorems is achieved by applying the general
results discussed in Section~\ref{s:mainris} and the
model dependent lemmas given above.
Indeed, 
Theorem~\ref{t:bc050} follows by
Theorem~\ref{t:addition01} and Lemmas~\ref{t:bc010}--\ref{t:bc040}, 
whereas 
Theorem~\ref{t:bc060} follows by
Theorem~\ref{t:addition02},
and Lemmas~\ref{t:bc010}--\ref{t:bc040}.



\subsection{Some more notation}
\label{s:def-modello}
\par\noindent
In this section we collect some definitions that will be used in the
proof of the Lemmas~\ref{t:bc010}--\ref{t:bc040}.
We let
\begin{equation}
\label{lambdo}
\Lambda^{s}(\sigma):=
 \{x\!\in\!\Lambda\!:\sigma(x)=s\}
\end{equation}
for any $\sigma\in\cc{X}$ with $s\in\{-1,0,+1 \}$.
Recall $L$ denotes the side length of the squared lattice $\Lambda$.
Let $x=(x_1,x_2)\in\Lambda$;
for $\ell_1,\ell_2$ positive integers we let
$R^x_{\ell_1,\ell_2}$ be the collection of the sites
$\big((x_1+n_1)\!\!\!\mod L,(x_2+n_2)\!\!\!\mod L\big)$
for $n_i=0,\dots,x_i+\ell_i-1$ where $i=1,2$.
Roughly speaking, $R^x_{\ell_1,\ell_2}$ is
the rectangle on the torus of side lengths $\ell_1$ and $\ell_2$
drawn starting from $x$ and moving in the positive direction along
the two coordinate axes.
For $\ell$ a positive integer we let $Q^x_\ell:=R^x_{\ell,\ell}$.

We denote with $\mathcal{R}_{\ell_1,\ell_2}$ the set of the configurations
$\sigma\in\cc{X}$ which are
\textit{rectangular droplet} of zeroes with side lengths
$\ell_1$ and $\ell_2$ in a sea of minus, with
$\ell_1,\ell_2$ integers such that
$2\le\ell_1,\ell_2\le L-1$. More precisely,
$\sigma\in\mathcal{R}_{\ell_1,\ell_2}$ if and only if
there exists $x\in\Lambda$ such that either
$\Lambda^0(\sigma)=R^x_{\ell_1,\ell_2}$
and the spins in the
complementary set $\Lambda\setminus\Lambda^0(\sigma)$
are negative.
Moreover, we denote with
$\mathcal{R}^x_{\ell_1,\ell_2}\subset\cc{R}_{\ell_1,\ell_2}$
the rectangular droplet of zeroes with side lengths
$\ell_1$ and $\ell_2$
in a see of minuses and with the lower--left corner in $x$.

\begin{figure}[t]
 \begin{picture}(100,90)(-60,15)
 \thinlines
 \setlength{\unitlength}{0.045cm}
 \put(60,15){\line(1,0){40}}
 \put(60,15){\line(0,1){45}}
 \put(60,60){\line(1,0){40}}
 \put(100,15){\line(0,1){10}}
 \put(100,60){\line(0,-1){25}}
 \put(100,25){\line(1,0){5}}
 \put(100,35){\line(1,0){5}}
 \put(105,25){\line(0,1){10}}
 \put(75,63){${\scriptstyle \ell}$}
 \put(53,35){${\scriptstyle m}$}
 \put(70,30){\begin{turn}{30}{\scriptsize zeroes}\end{turn}}
 \put(105,50){\begin{turn}{-30}{\scriptsize minuses}\end{turn}}
 \put(160,15){\line(1,0){10}}
 \put(170,15){\line(0,-1){5}}
 \put(170,10){\line(1,0){15}}
 \put(185,10){\line(0,1){5}}
 \put(185,15){\line(1,0){15}}
 \put(160,15){\line(0,1){45}}
 \put(160,60){\line(1,0){40}}
 \put(200,15){\line(0,1){45}}
 \put(200,60){\line(0,-1){30}}
 \put(175,63){${\scriptstyle \ell}$}
 \put(153,35){${\scriptstyle m}$}
 \put(170,30){\begin{turn}{30}{\scriptsize zeroes}\end{turn}}
 \put(205,50){\begin{turn}{-30}{\scriptsize minuses}\end{turn}}
 \end{picture}
 \caption{Schematic representation of the configurations
          $\mathcal{C}_{\ell,m}^x(2;E)$ (on the left) and $\mathcal{C}_{\ell,m}^x(3;S)$ (on the right).
          }
 \label{fig:fig12}
\end{figure}
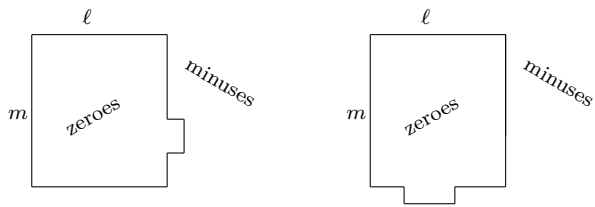

Given a rectangular droplet in $\cc{R}^x_{\ell_1,\ell_2}$,
we let
$N,E,S$, and $W$ represent respectively the north, east, south,
and west side
of the rectangular droplet.
For $D\in\{N,E,S,W\}$,
we denote by $\mathcal{C}_{\ell_1,\ell_2}^x(n;D)$
the configuration obtained by adding
a zero protuberance of length $n$ to the $D$--side
of the rectangular droplet
(see Fig.~\ref{fig:fig12}).
Note that $n$ is a not negative integer bounded by
$\ell_1$ if $D\in\{N,S\}$
and
$\ell_2$ if $D\in\{W,E\}$.
Note, also, that
$\cc{C}^x_{\ell_1,\ell_2}(0;D)=\cc{R}^x_{\ell_1,\ell_2}$,
$\cc{C}^x_{\ell_1,\ell_2}(\ell_1;D)=\cc{R}^x_{\ell_1,\ell_2+1}$
if $D\in\{N,S\}$,
and
$\cc{C}^x_{\ell_1,\ell_2}(\ell_2;D)=\cc{R}^x_{\ell_1+1,\ell_2}$
if $D\in\{E,W\}$.

Moreover, we denote  with $\mathcal{C}_{\ell_1,\ell_2}^x(n)$ the
configuration obtained by adding
a zero protuberance of length $n$ to any of the four side of the
rectangular droplet.
(i.e., $\mathcal{C}_{\ell_1,\ell_2}^x(n)
=\bigcup_D \mathcal{C}_{\ell_1,\ell_2}^x(n;D)$).
We also let\\
$\mathcal{C}_{\ell_1,\ell_2}(n)
=\bigcup_{x\in\Lambda} \mathcal{C}_{\ell_1,\ell_2}^x(n)$.

Given a configuration in
$\mathcal{C}_{\ell_1,\ell_2}^x(n;D)$
its
\emph{rectangular envelope}
is the configuration obtained by flipping to zero the minuses on the
side occupied by the protuberance.
Then, we have that
the rectangular envelope of a configuration in
$\mathcal{C}_{\ell_1,\ell_2}^x(n;D)$ belongs to
either
$\mathcal{R}_{\ell_1+1,\ell_2}$
or
$\mathcal{R}_{\ell_1,\ell_2+1}$
depending
if $D\in\{E,W\}$ or $D\in\{N,S\}$.

Given a configuration in
$\sigma_0\in\mathcal{C}_{\ell_1,\ell_2}(n;D)$
with $n\ge1$
a \emph{standard growing path} is a path
$(\sigma_0,\sigma_1,\dots,\sigma_k)$
such that $\sigma_{i+1}$ is obtained by enlarging
by one zero spin the protuberance in $\sigma_i$ and $\sigma_k$
is the rectangular envelope of $\sigma_0$.
Note that $k=\ell_1-n$ and $k=\ell_2-n$
if $D\in\{E,W\}$ or $D\in\{N,S\}$, respectively.

Given a configuration in
$\sigma_0\in\mathcal{C}_{\ell_1,\ell_2}(n;D)$
with $n\ge1$
a \emph{standard shrinking path} is a path
$(\sigma_0,\sigma_1,\dots,\sigma_k)$
such that $\sigma_{i+1}$ is obtained by flipping to minus
one of the zero spins of the protuberance having at most
two neighboring minuses
and $\sigma_k$ is the configuration obtained by flipping to minus
all the spin in the protuberance of $\sigma_0$.

In case of a stripe winding around the torus, i.e.,  
$\ell_1\vee\ell_2=L$, we use the
 same notation adopted for the rectangular droplets:  $\mathcal{R}_{\ell_1,\ell_2}$  is the set of the
$\sigma\in\cc{X}$ which are either
\textit{horizontal stripes} of zeroes in a see of minus if $\ell_1=L$ or  \textit{vertical stripes} if  $\ell_2=L$.
Moreover, we denote  with $\mathcal{C}_{\ell_1,\ell_2}^x(n)$ the
configuration obtained by adding
a zero protuberance of length $n$ to any of the two sides of the
stripe with length smaller than $L$.

For any $s\in\{-1,0,+1\}$ and $x\in\Lambda$,
we define the spin--flip operator $S^x_{s}:\cc{X}\to\cc{X}$ by letting
$S^x_s\sigma$ be the configuration such that
\begin{equation}
\label{d:soperator}
S^x_{s}\sigma(y):=
\left\{
\begin{array}{ll}
s& \textnormal{ for } y=x, s\neq \sigma(x)\\
\sigma(y) & \textnormal{ otherwise }
\end{array}
\right.
\end{equation}

Recall the definition of path in the configuration space given
just above \eqref{height}.
A path $\omega=(\omega_1,\dots,\omega_m)\in\cc{X}^m$ is
\emph{downhill} if and only if
$H(\omega_i)\ge H(\omega_{i+1})$ for any $i=1,\dots,m-1$.
We say that
a configuration $\sigma\in\cc{X}$
is a \emph{local minimum} of the Hamiltonian if and only if
$H(\eta)\ge H(\sigma)$ for any $\eta\in\cc{X}$ such that $(\sigma,\eta)\in Q$.
For our purposes
it is useful to introduce the notion of
\emph{strict downhill} path by saying that a path
$\omega=(\omega_1,\dots,\omega_m)\in\cc{X}^m$ is
\emph{strict downhill} if and only if
$H(\omega_i)> H(\omega_{i+1})$ for any $i=1,\dots,m-1$
and $\sigma_m$ is a local minimum of the Hamiltonian.

\subsection{Proof of the lemmas concerning the Blume--Capel model}
\label{s:proof-esempio}
\par\noindent
\textit{Proof of Lemma~\ref{t:bc010}.\/}
By \cite[Theorem~4.10]{CN2013} $\cc{X}_{\textrm{m}}=\{\muno,\zero\}$ and
$\Gamma_\textrm{m}=\Gamma_\textrm{c}$.
By direct inspection of the Hamiltonian
\eqref{cap:002} it follows that
$\cc{X}_\textrm{s}=\{\puno\}$ and $H(\puno)>H(\zero)$ (see, also,
the comments below \cite[Condition~4.7]{CN2013}).
\qed

The proof of Lemma~\ref{t:bc020} needs the discussion of some
preliminary results aimed to describe the paths followed by the
system when it performs the transition from $\muno$ to $\zero$.
The first step is that of computing energy differences
between configurations differing for a single spin.
Such a difference will depend only on the configuration
in a cross--shaped neighborhood centered at the site with
differing spins. Thus,
for any $x\in\Lambda$, we denote with
$V(x)$ the neighborhood of $x$ defined as
$V(x)=\{y\in \Lambda: d(x,y)\leq 1\}$, where $d(\cdot,\cdot)$ is the
Euclidean distance on the torus.
Given a configuration $\sigma$,
the effect on the Hamiltonian of a change of the spin at site
$x$ will depend only on the configuration $\sigma_{V(x)}$
obtained by restricting $\sigma$ to $V(x)$ (see, the definition
of restriction given above \eqref{lc}). All the possible
cases are summarized in the Table~\ref{tab:main},
where the configurations $A_1,A_2,A_3,B_1,\dots,O_3$ are listed
and the corresponding difference of energies are reported.

\begin{table}
\label{tab:main}
\centering
\scalebox{.80}{
\begin{tabular}{|c|c|c|c|c|c|c|}
\hline
\phantom{A}&
$1$
& $2$
& $3$
& $H(2)-H(1)$
& $H(3)-H(1)$
& $H(3)-H(2)$
\\
\hline
A&
\begin{tiny}
\begin{tabular}{ccc}
&$-$&\\
$-$&$-$&$-$\\
&$-$&
\end{tabular}
\end{tiny}
&
\begin{tiny}
\begin{tabular}{ccc}
&$-$&\\
$-$&$0$&$-$\\
&$-$&
\end{tabular}
\end{tiny}
&
\begin{tiny}
\begin{tabular}{ccc}
&$-$&\\
$-$&$+$&$-$\\
&$-$&
\end{tabular}
\end{tiny}
& $4-h$
& $16-2h$
& $12-h$
\\
\hline
B&
\begin{tiny}
\begin{tabular}{ccc}
&$-$&\\
$-$&$-$&$-$\\
&$0$&
\end{tabular}
\end{tiny}
&
\begin{tiny}
\begin{tabular}{ccc}
&$-$&\\
$-$&$0$&$-$\\
&$0$&
\end{tabular}
\end{tiny}
&
\begin{tiny}
\begin{tabular}{ccc}
&$-$&\\
$-$&$+$&$-$\\
&$0$&
\end{tabular}
\end{tiny}
&
$2-h$
&
$12-2h$
&
$10-h$
\\
\hline

C&
\begin{tiny}
\begin{tabular}{ccc}
&$-$&\\
$-$&$-$&$-$\\
&$+$&
\end{tabular}
\end{tiny}
&
\begin{tiny}
\begin{tabular}{ccc}
&$-$&\\
$-$&$0$&$-$\\
&$+$&
\end{tabular}
\end{tiny}
&
\begin{tiny}
\begin{tabular}{ccc}
&$-$&\\
$-$&$+$&$-$\\
&$+$&
\end{tabular}
\end{tiny}
&
$-h$
&
$8-2h$
&
$8-h$
\\
\hline

D&
\begin{tiny}
\begin{tabular}{ccc}
&$-$&\\
$0$&$-$&$-$\\
&$0$&
\end{tabular}
\end{tiny}
&
\begin{tiny}
\begin{tabular}{ccc}
&$-$&\\
$0$&$0$&$-$\\
&$0$&
\end{tabular}
\end{tiny}
&
\begin{tiny}
\begin{tabular}{ccc}
&$-$&\\
$0$&$+$&$-$\\
&$0$&
\end{tabular}
\end{tiny}
&
$-h$
&
$8-2h$
&
$8-h$
\\
\hline

E&
\begin{tiny}
\begin{tabular}{ccc}
&$-$&\\
$+$&$-$&$-$\\
&$0$&
\end{tabular}
\end{tiny}
&
\begin{tiny}
\begin{tabular}{ccc}
&$-$&\\
$+$&$0$&$-$\\
&$0$&
\end{tabular}
\end{tiny}
&
\begin{tiny}
\begin{tabular}{ccc}
&$-$&\\
$+$&$+$&$-$\\
&$0$&
\end{tabular}
\end{tiny}
&
$-2-h$
&
$4-2h$
&
$6-h$
\\
\hline
F&
\begin{tiny}
\begin{tabular}{ccc}
&$-$&\\
$+$&$-$&$-$\\
&$+$&
\end{tabular}
\end{tiny}
&
\begin{tiny}
\begin{tabular}{ccc}
&$-$&\\
$+$&$0$&$-$\\
&$+$&
\end{tabular}
\end{tiny}
&
\begin{tiny}
\begin{tabular}{ccc}
&$-$&\\
$+$&$+$&$-$\\
&$+$&
\end{tabular}
\end{tiny}
&
$-4-h$
&
$-2h$
&
$4-h$
\\
\hline
G&
\begin{tiny}
\begin{tabular}{ccc}
&$-$&\\
$0$&$-$&$0$\\
&$0$&
\end{tabular}
\end{tiny}
&
\begin{tiny}
\begin{tabular}{ccc}
&$-$&\\
$0$&$0$&$0$\\
&$0$&
\end{tabular}
\end{tiny}
&
\begin{tiny}
\begin{tabular}{ccc}
&$-$&\\
$0$&$+$&$0$\\
&$0$&
\end{tabular}
\end{tiny}
&
$-2-h$
&
$4-2h$
&
$6-h$
\\
\hline
H&
\begin{tiny}
\begin{tabular}{ccc}
&$-$&\\
$0$&$-$&$+$\\
&$0$&
\end{tabular}
\end{tiny}
&
\begin{tiny}
\begin{tabular}{ccc}
&$-$&\\
$0$&$0$&$+$\\
&$0$&
\end{tabular}
\end{tiny}
&
\begin{tiny}
\begin{tabular}{ccc}
&$-$&\\
$0$&$+$&$+$\\
&$0$&
\end{tabular}
\end{tiny}
&
$-4-h$
&
$-2h$
&
$4-h$
\\
\hline
J&
\begin{tiny}
\begin{tabular}{ccc}
&$-$&\\
$0$&$-$&$+$\\
&$+$&
\end{tabular}
\end{tiny}
&
\begin{tiny}
\begin{tabular}{ccc}
&$-$&\\
$0$&$0$&$+$\\
&$+$&
\end{tabular}
\end{tiny}
&
\begin{tiny}
\begin{tabular}{ccc}
&$-$&\\
$0$&$+$&$+$\\
&$+$&
\end{tabular}
\end{tiny}
&
$-6-h$
&
$-4-2h$
&
$2-h$
\\
\hline
K&
\begin{tiny}
\begin{tabular}{ccc}
&$-$&\\
$+$&$-$&$+$\\
&$+$&
\end{tabular}
\end{tiny}
&
\begin{tiny}
\begin{tabular}{ccc}
&$-$&\\
$+$&$0$&$+$\\
&$+$&
\end{tabular}
\end{tiny}
&
\begin{tiny}
\begin{tabular}{ccc}
&$-$&\\
$+$&$+$&$+$\\
&$+$&
\end{tabular}
\end{tiny}
&
$-8-h$
&
$-8-2h$
&
$-h$
\\
\hline
I&
\begin{tiny}
\begin{tabular}{ccc}
&$0$&\\
$0$&$-$&$0$\\
&$0$&
\end{tabular}
\end{tiny}
&
\begin{tiny}
\begin{tabular}{ccc}
&$0$&\\
$0$&$0$&$0$\\
&$0$&
\end{tabular}
\end{tiny}
&
\begin{tiny}
\begin{tabular}{ccc}
&$0$&\\
$0$&$+$&$0$\\
&$0$&
\end{tabular}
\end{tiny}
&
$-4-h$
&
$-2h$
&
$4-h$
\\
\hline
L&
\begin{tiny}
\begin{tabular}{ccc}
&$0$&\\
$0$&$-$&$0$\\
&$+$&
\end{tabular}
\end{tiny}
&
\begin{tiny}
\begin{tabular}{ccc}
&$0$&\\
$0$&$0$&$0$\\
&$+$&
\end{tabular}
\end{tiny}
&
\begin{tiny}
\begin{tabular}{ccc}
&$0$&\\
$0$&$+$&$0$\\
&$+$&
\end{tabular}
\end{tiny}
&
$-6-h$
&
$-4-2h$
&
$2-h$
\\
\hline
M&
\begin{tiny}
\begin{tabular}{ccc}
&$0$&\\
$+$&$-$&$0$\\
&$+$&
\end{tabular}
\end{tiny}
&
\begin{tiny}
\begin{tabular}{ccc}
&$0$&\\
$+$&$0$&$0$\\
&$+$&
\end{tabular}
\end{tiny}
&
\begin{tiny}
\begin{tabular}{ccc}
&$0$&\\
$+$&$+$&$0$\\
&$+$&
\end{tabular}
\end{tiny}
&
$-8-h$
&
$-8-2h$
&
$-h$
\\
\hline
N&
\begin{tiny}
\begin{tabular}{ccc}
&$0$&\\
$+$&$-$&$+$\\
&$+$&
\end{tabular}
\end{tiny}
&
\begin{tiny}
\begin{tabular}{ccc}
&$0$&\\
$+$&$0$&$+$\\
&$+$&
\end{tabular}
\end{tiny}
&
\begin{tiny}
\begin{tabular}{ccc}
&$0$&\\
$+$&$+$&$+$\\
&$+$&
\end{tabular}
\end{tiny}
&
$-10-h$
&
$-12-2h$
&
$2-h$
\\
\hline
O&
\begin{tiny}
\begin{tabular}{ccc}
&$+$&\\
$+$&$-$&$+$\\
&$+$&
\end{tabular}
\end{tiny}
&
\begin{tiny}
\begin{tabular}{ccc}
&$+$&\\
$+$&$0$&$+$\\
&$+$&
\end{tabular}
\end{tiny}
&
\begin{tiny}
\begin{tabular}{ccc}
&$+$&\\
$+$&$+$&$+$\\
&$+$&
\end{tabular}
\end{tiny}
&
$-12-h$
&
$-16-2h$
&
$-4-h$
\\
\hline
\end{tabular}}
\caption{The first four columns define the configurations
$A_i,\dots,O_i$ with $i=1,2,3$. In the configurations
$-$ and $+$ denote $-1$ and $+1$, respectively.
The column $H(i)-H(j)$ reports the energy difference between
the configurations depicted in the columns $i$ and $j$.
}
\end{table}

We state and prove now the following Lemma regarding enlarging
a protuberance of a rectangular droplet following the energy drift.
\begin{lemma}
\label{l:allprot}
Given $\sigma\in \mathcal{C}_{\ell_1,\ell_2}(1)$,
any strict downhill path started at $\sigma$ is either
a standard shrinking or growing path.
\end{lemma}

\par\noindent
\textit{Proof.\/}
Let us start considering the case $\ell_1\vee\ell_2<L$, i.e.,
proper rectangular droplets.
Given $\sigma\in \mathcal{C}_{\ell_1,\ell_2}(1)$,
by table~\ref{tab:main} we deduce that for any $y\in\Lambda$,
the restriction of $\sigma$ to the cross neighbor $V(y)$  is such that
\begin{equation}
\label{e:vtype}
\sigma_{V(y)}\in\{A_1,B_1,B_2,D_1,D_2,G_2,I_2\}
\end{equation}
For any $i\in\Lambda$ we denote with $H_i(\eta)$,
the contribution of the site $i$ to the energy $H(\eta)$, i.e., 
$H_i(\eta):=\sum_{j:|j-i|=1}(\eta(i)-\eta(j))^2-h\sigma(i)$,
where the sum is over the nearest--neighbor sites of $i$.
Consider, now, a strict downhill path $(\omega_1,\dots,\omega_k)$
started at $\sigma$.
Since in a strict downhill path at each step there is one single spin
which is flipped, we have that
for any $m\in\{1,\ldots,k-1\}$ there exists a site
$i_m\in\Lambda$
and $s_m\in\{-1,0,+1\}\setminus\{\omega_m(i_m)\}$
such that
$$
H(\omega_{m+1})-H(\omega_m)
=H_{i_m}(S^{i_m}_{s_m} {\omega_m})
 -H_{i_m}({\omega_m})
$$
By (\ref{e:vtype}) we have that
$(\omega_1)_{V(x_1)}\in\{A_1,B_1,B_2,D_1,D_2,G_2,I_2\}$
and, by Table~\ref{tab:main}, it follows immediately that
\begin{equation}
\label{e:cases}
H(\omega_{2})-H(\omega_1)
\left\{
\begin{array}{ll}
=-2+h&\textnormal{if } (\omega_1)_{V(x_1)}=B_2 \textrm{ and } s_1=-1,\\
=-h&\textnormal{if } (\omega_1)_{V(x_1)}=D_1 \textrm{ and } s_1=0,\\
>0&\textrm{otherwise.}
\end{array}
\right.
\end{equation}

Therefore, a strict downhill path will either
remove the protuberance (i.e., $(\omega_1)_{V(x_1)}=B_2$ and $s_1=-1$)
or enlarge by one zero the existing one
(i.e., $(\omega_1)_{V(x_1)}=D_1$ and $s_1=0$).

In the first case
$\omega_2\in \mathcal{R}_{\ell_1,\ell_2}$.
Since
$(\omega_2)_{V(y)}\in\{A_1,B_1,D_2,G_2,I_2\}$ for any $y\in\Lambda$,
by Table~\ref{tab:main} it follows that $\omega_2$ is a local
minimum of the Hamiltonian. Hence, in this first case, the
standard shrinking sequence is found.

In the second case, i.e., when the protuberance is enlarged,
for  any $y\in \Lambda$,
$(\omega_2)_{V(y)}\in \{A_1,B_1,D_1,D_2,G_2,I_2\}$.
Notice that there are not anymore neighbors of type
$B_2$, because the protuberance, now, is at least wide two sites.
Hence, if we use the same argument used for $\omega_1$,
we deduce that the only way of lowering the energy is by
enlarging the protuberance along the side.
We can repeat the same argument until all the line is filled.
In this way the standard growing sequence is found.

In case of stripes, i.e., $\ell_1\vee\ell_2=L$,  the proof is almost identical, with only minor adjustments: for instance in the set $(\ref{e:vtype})$ is
not present anymore the neighborhood   $D_2$.
\qed

Recall the definition of external boundary given in Section~\ref{s:rel}
and note that in the Blume--Capel case the external boundary of a
subset of the
configuration space is made of all those configurations not belonging
to such a set and such that by changing the value of one single spin
the configuration that is obtained belongs to the set.
We call a nonempty set $C\subset\mathcal{S}$ a
\emph{cycle} if it is either a singleton or a connected set such that
\begin{equation}
\label{e:nontrivialcycle}
\max_{x\in C} H(x)<H(F(\partial C))
\end{equation}
where
\begin{displaymath}
F(\partial C):=\textnormal{argmin}_{y\in \partial C} H(y)
\end{displaymath}
A non-trivial cycle is a cycle for which (\ref{e:nontrivialcycle}) holds.

Following \cite{NZB}, we define the \emph{principal boundary}
$\mathcal{B}(C)$ of a cycle $C$, as
\begin{displaymath}
\mathcal{B}(C):=
\left\{
\begin{array}{ll}
F(\partial C)& \textnormal{ if } C \textnormal{ is a non--trivial cycle},\\
\{z\in\partial C:\, H(z)\leq H(y)\}&  \textnormal{ if } C=\{y\} \textnormal{ is a trivial cycle}
\end{array}
\right.
\end{displaymath}

For any rectangular droplet or stripe $\zeta\in\mathcal{R}_{\ell_1,\ell_2}$
we define the cycle
\begin{equation}
\label{e:rcycle}
\mathcal{A}_{\zeta}=\{\sigma\in\mathcal{X}: \Phi(\zeta,\sigma)<H(\zeta)+2-h\}
\end{equation}
made of all the configurations that can be reached starting from $\zeta$
via a path whose energy stays below $H(\zeta)+2-h$.
The following lemma gives a precise characterization of the
minima of the energy of the external boundary of the cycle
$\mathcal{A}_\zeta$ for any $\zeta\in\cc{R}_{\ell_1,\ell_2}$
with $\ell_1,\ell_2\ge \ell_\textrm{c}$, see \eqref{lc}.

\begin{lemma}
\label{l:rett0}
For any $\zeta\in\cc{R}_{\ell_1,\ell_2}$
such that
$\ell_1,\ell_2\ge \ell_\textrm{c}$, it holds:
$i)$
$
\mathcal{B}(\mathcal{A}_\zeta)
=\{\eta\in \mathcal{C}_{\ell_1,\ell_2}(1):
\exists j\in \Lambda \,\,\textrm{ such that }\,\, \eta=S_{0}^j \zeta\}
$;
$ii)$
$F(\mathcal{A}_{\zeta})=\zeta$.
\end{lemma}
\par\noindent
\textit{Proof.\/}
Item $i)$.
Let us start considering the case $\ell_1\vee\ell_2<L$, i.e., a proper
rectangular droplet.
Let $\zeta\in\mathcal{R}_{\ell_1,\ell_2}$.
As we have noted in the proof of Lemma~\ref{l:allprot}
we have that
${\zeta}_{V(i)}\in \{A_1,B_1,D_2,G_2,I_2\}$, for any $i\in \Lambda$. This implies that (see
Table~\ref{tab:main})
the rectangular droplet  $\zeta$ is a local minimum of the Hamiltonian.
Moreover, by using the results in the table one has that
\begin{displaymath}
H(S^{i}_{s} \zeta)-H(\zeta)
\ge
2-h
\end{displaymath}
for any $i\in\Lambda$ excepted
for the case of the
corner erosion, namely, for $i$ equal to one of the four sites such
that $\zeta_{V(i)}=D_2$ and $s=-1$. In these cases $H_{i}(S^{i}_{s} \zeta)-H_{i}(\zeta)=h$.
Hence, all the flips but the corner erosion
yield a configuration outside $\cc{A}_\zeta$. In particular, we remark that the equality $2-h$ is attained by adding a protuberance to the rectangular configuration, i.e., $S^{i}_{0} \zeta\in \mathcal{C}_{\ell_1,\ell_2}(1)$. 
This implies that
$\mathcal{C}_{\ell_1,\ell_2}(1)\subseteq \textnormal{argmin}_{\eta\in\partial \mathcal{A}_{\zeta}}H(\eta)$.

Let us give the following definitions. Given a configuration $\eta$, we define the $0$--\emph{rectangular envelope} the configuration $R(\eta)$ such that $\Lambda^{0}(R(\eta))$ is the smallest rectangle containing $\Lambda^{0}(\eta)$, where $\Lambda^0(\cdot)$ is defined in (\ref{lambdo}).
Furthermore, we will call \emph{corner--erosion} the spin flip from  $0$ to $-1$ in the site $i$ with neighborhood $V(i)$ of type $D_2$.
Moreover, we define the set of configurations:
\begin{displaymath}
\Xi_1:=\{\sigma\in\mathcal{S}: \exists i\in \Lambda\, \textnormal{ s.t }  \zeta_{V(i)}=D_2,
\sigma= S_{-1}^i(\zeta), \, R(\sigma)=\zeta\}
\end{displaymath}
in words, $\Xi_1$ is the set of all the configurations obtained from $\zeta$ 
by one step of {corner--erosion}.
We note that for any configuration 
$\sigma\in \Xi_1$, $i\in \Lambda$  we have that
${\sigma}_{V(i)}\in \{A_1,B_1,D_1,D_2,G_2,I_2\}$.
By inspection of  Table~\ref{tab:main}, it follows  that if ${\sigma}_{V(i)}\in \{A_1,B_1,D_1,G_2,I_2\}$, then 
\begin{eqnarray*}
H(S^{i}_{s} \sigma)-H(\zeta)&=&H(S^{i}_{s} \sigma)-H(\sigma)+H(\sigma)-H(\zeta)\\&=&H(S^{i}_{s} \sigma)-H(\sigma)+h
\ge
2-h+h=2
\end{eqnarray*}
so that  $S^{i}_{s} \sigma\notin \mathcal{A}_\zeta$ and $S^{i}_{s}\sigma \notin\textnormal{argmin}_{\eta\in\partial \mathcal{A}_{\zeta}}H(\eta)$. Otherwise, if ${\sigma}_{V(i)}=D_2$ we have that $H(S^{i}_{s} \sigma)-H(\zeta)>2-h$ for $s=+1$, while
for $s=-1$ we have:
\begin{displaymath}
H(S^{i}_{-1} \sigma)-H(\zeta)=2h
\end{displaymath}
and $S^{i}_{-1} \sigma\in\mathcal{A}_\zeta$, that gives a second corner--erosion.
This suggests the following definition, for any $k\geq 2$:
\begin{displaymath}
\Xi_k:=\{\eta\in\mathcal{S}:  \exists i\in \Lambda,\, \exists \gamma\in \Xi_{k-1} \textnormal{ s.t. }  \gamma_{V(i)}=D_2,
\eta= S_{-1}^i(\gamma), \, R(\eta)=\zeta\}
\end{displaymath}
In words  $\Xi_k$ is the set of all the configurations obtained by $\zeta$  with $k$ corner--erosions.  Notice that $S^{i}_{-1} \sigma\in \Xi_2$.

We show now that if $k\leq \ell_c-2$ then $\Xi_k\subseteq \mathcal{A}_\zeta$.
If $\gamma\in \Xi_k$, then we have indeed that
$H(\gamma)=k\,h+H(\zeta)<(2/h+1-2)h+H(\zeta)=2-h+H(\zeta)$, where we used  $\ell_\rr{c}:=\lfloor\frac{2}{h}\rfloor+1$.

If $k\leq \ell_c-3$,  for any $\gamma\in \Xi_k$, $i\in \Lambda$  we have that:\\
 ${\gamma}_{V(i)}\in \{A_1,B_1,D_1,D_2,G_2,I_2\}$.
Again, by inspection of  Table~\ref{tab:main}, it follows  that,  for   $i\in\Lambda$ such that $\gamma_{V(i)}\in \{A_1,B_1,D_1,G_2,I_2\}$:

$$
H(S^{i}_{s} \gamma)-H(\zeta)
>2-h
$$
and $S^{i}_{s}(\gamma)\in\partial A_{\zeta}\setminus \textnormal{argmin}_{ \eta\in\partial \mathcal{A}_{\zeta}}H(\eta)$.
If $i$ is such that $\gamma_{V(i)}= D_2$, we have that $H(S^{i}_{s} \gamma)-H(\zeta)>2-h$ for $s=+1$, while
for $s=-1$ we have:

$$
H(S^{i}_{-1} \gamma)-H(\zeta)=h(k+1)<2-h
$$
so that $S^{i}_{-1}(\gamma)\in A_{\zeta}$. In words, from $\gamma\in\Xi_k$ any spin--flip that is not corner--erosion increses the energy by more than $2-h$.
Using similar arguments as above,  for any $\gamma\in\Xi_{\ell_c-2}$ and  for any  $i\in\Lambda$ we have that  ${\gamma}_{V(i)}\in \{A_1,B_1,D_1,D_2,G_2,I_2\}$. If $\gamma_{V(i)}\in \{A_1,B_1,D_1,G_2,I_2\}$, we get:
$$
H(S^{i}_{s} \gamma)-H(\zeta)
>2-h
$$
so that $S^{i}_{s}(\gamma)\in\partial A_{\zeta}\setminus \textnormal{argmin}_{ \eta\in\partial \mathcal{A}_{\zeta}}H(\eta)$.
If $i$  is such that $\gamma_{V(i)}= D_2$, it follows:
$$
H(S^{i}_{-1} \gamma)-H(\zeta)=h(\ell_c-1)>2-h
$$
so that $S^{i}_{-1}(\gamma)\in\partial A_{\zeta}\setminus \textnormal{argmin}_{ \eta\in\partial \mathcal{A}_{\zeta}}H(\eta)$.
In words, from $\gamma\in\Xi_{\ell_c-2}$ any spin--flip, including the corner--erosion,  increases the energy by more than $2-h$.
This conclude the proof of item $i)$, since we proved that $\mathcal{C}_{\ell_1,\ell_2}(1)\subseteq \textnormal{argmin}_
{\eta\in\partial\mathcal{A}_\zeta} H(\eta)$  and for any other path the configuration reached when exiting $\mathcal{A}_{\zeta}$ does not belong to  $\textnormal{argmin}_
{\eta\in\partial\mathcal{A}_\zeta} H(\eta)$.

\smallskip
Item $ii)$.\
Starting from $\zeta$ by $\ell_c-2$ corner--erosions, it is not possible to change the rectangular envelope, since $\ell_1$ and $\ell_2$ are not smaller than $\ell_c$. Furthermore,  for any $\gamma\in\Xi_k$, with $k\leq \ell_c-2$, every sites belonging to the rectangular envelope  $\Lambda^0(R(\gamma))$ is such that there are at least other two neighboring sites with spin zero.
By the proof of item $i)$, we have that starting from $\zeta$, all the possible spin flipped configurations $S_{s}^j\zeta$ for  $j\in\Lambda$ and $s\in\{-1,0,+1\}$, belongs either to $\Xi_1$ or to $\mathcal{A}_\zeta^\emph{c}$.
Furthermore, for any $\eta\in \Xi_1$, all the possible spin flipped configurations $S_{s}^i\eta$ for  $i\in\Lambda$ and $s\in\{-1,0,+1\}$, belongs either to $\Xi_2\cup \{\zeta\}$ or  to  $\mathcal{A}_\zeta^c$.
 For any $\eta\in \Xi_k$, with $k\leq \ell_c-3$, all the possible spin flipped configurations $S_{s}^i\eta$, belongs either to $\Xi_{k+1}\cup \Xi_{k+2}$ or  to  $\mathcal{A}_\zeta^c$.
 Finally, for any $\eta\in \Xi_{\ell_c-2}$,  all the possible spin flipped configurations $S_{s}^j\eta$, belongs either to $\Xi_{\ell_c-3}$ or  to  $\mathcal{A}_\zeta^c$.
 Putting together the previous statements, we can analyze all the paths joining $\zeta$ to $\eta$ with $\eta\in \mathcal{A}_\zeta$ and we get that $\omega_{k^\prime}\in \bigcup_{k=1}^{\ell_c-2} \Xi_k\cup \zeta$, with $k^\prime\leq \ell_c-2$.
Therefore in any line and row of $\Lambda^0(\eta)$ there are at lest two sites with spin $0$. Thus, we have that
\begin{displaymath}
\mathcal{A}_\zeta=\bigcup_{k=1}^{\ell_c-2} \Xi_k\cup \zeta
\end{displaymath}
Since $H(\Xi_k)>H(\zeta)$, for $k\leq \ell_c-2$,
the item follows.

In case of stripes, i.e., $\ell_1\vee\ell_2=L$,  the proof is simpler. In fact it is not possible anymore to erode a corner:
we have that
${\zeta}_{V(i)}\in \{A_1,B_1,G_2,I_2\}$, for any $i\in \Lambda$. This implies that (see
Table~\ref{tab:main})
\begin{displaymath}
H(S^{i}_{s} \zeta)-H(\zeta)
\ge
2-h
\end{displaymath}
Hence, all the flips
yield a configuration outside $\cc{A}_\zeta$.
\qed

\begin{lemma}
\label{l:rett1}
For any $\zeta\in\cc{R}_{\ell_\textrm{c}-1,\ell_\textrm{c}+1}$,
it holds:
$i)$
$\mathcal{B}(\mathcal{A}_\zeta)$ is made by the configurations
in $\mathcal{C}_{\ell_\textrm{c}-1,\ell_\textrm{c}}(1)$ obtained
by flipping to zero all the spins but one on one of the
two shortest side of the zero droplet in $\zeta$;
$ii)$
$F(\mathcal{A}_{\zeta})=\zeta$.
\end{lemma}
\par\noindent
\textit{Proof.\/}
The Lemma can be proved by following the same strategy used in
the proof of Lemma~\ref{l:rett0}.
\qed

\begin{figure}[t]
\centering
\includegraphics[width=9cm,height=6cm]{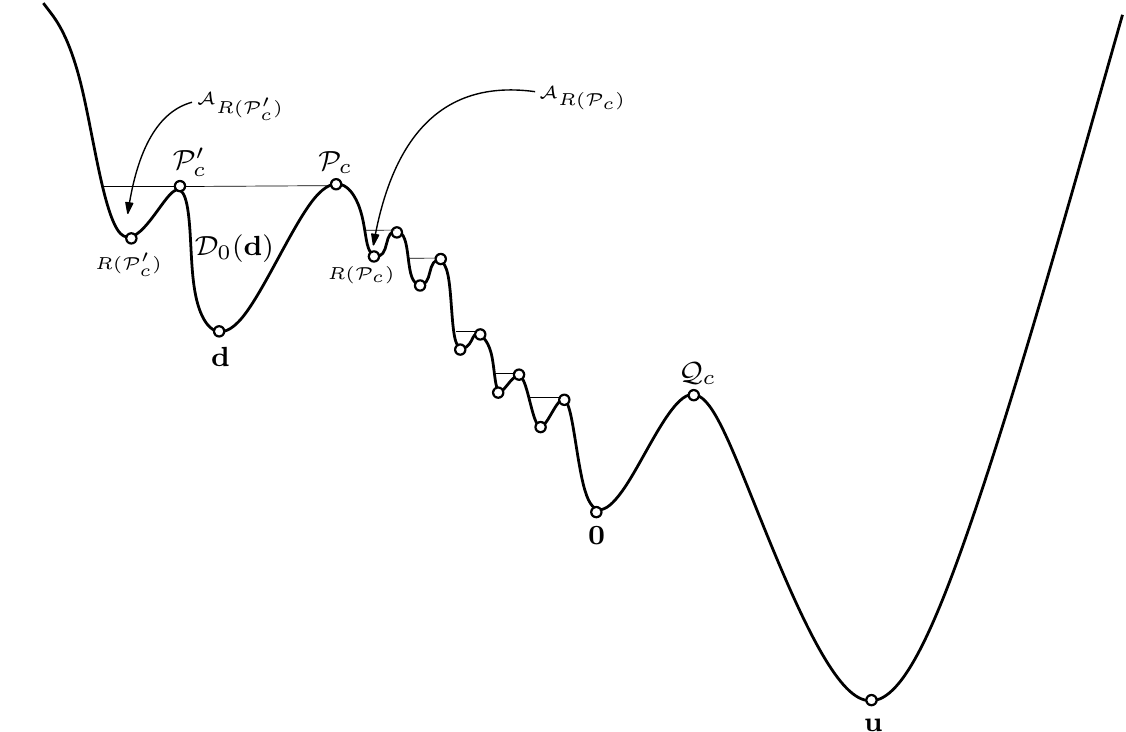}
\caption{Energy landscape and construction of $\mathcal{T}_\zero(\muno)$}
\label{fig:landscape}
\end{figure}


We finally come to the proof of  Lemma~\ref{t:bc020}. Our strategy is similar
to the one used in [Lemma 3.13, \cite{NZB}].
We let $\mathcal{T}_{\zero}(\muno)$ be the collection of the
following trivial and non--trivial pairwise disjoint cycles
(see Fig.~\ref{fig:landscape}):
\begin{enumerate}
\item\label{i01}
$\mathcal{D}_\zero(\muno)
:=\{\eta\in\mathcal{X}:\,\Phi(\eta,\muno)-H(\muno)<\Gamma_\textrm{c}\}$,
\item\label{i02}
all the configurations in $\mathcal{P}_c$,
\item\label{i03}
the non--trivial cycles
$\mathcal{A}_\zeta$ for
$\zeta\in\mathcal{R}_{\ell_1,\ell_2}$ such that
$\min \{\ell_1,\ell_2\}\geq\ell_c$
and
$\max \{\ell_1,\ell_2\}\leq L-2$;
\item\label{i04}
any configuration obtained by adding a
$0$ protuberance of length $\ell$ to one side of
$\zeta\in\mathcal{R}_{\ell_1,\ell_2}$ with
$\min \{\ell_1,\ell_2\}\geq \ell_c $
and
$\max \{\ell_1,\ell_2\}\leq L-3$,
and such that $\ell_1-\ell \geq \ell_c$
(respectively $\ell_2-\ell \geq \ell_c$) if the protuberance
has been added to the side with length $\ell_1$ (respectively $\ell_2$);
\item\label{i05}
any configuration
obtained by flipping to zero the spins
associated with any nearest neighbor connected\footnote{A subset
$A$ of the torus $\Lambda$ is said to be nearest neighbor connected if and
only if for any $x,y\in A$ there exists a sequence of pairwise
nearest neighbor sites of $A$ such that the first is $x$ and the last
is $y$.}
subset of the set of sites adjacent to the
shortest sides of any rectangular droplet $\zeta\in\mathcal{R}_{\ell_1,\ell_2}$
such that
$\ell_\textrm{c}\le\min\{\ell_1,\ell_2\}\le L-3$
and $\max\{\ell_1,\ell_2\}= L-2$;
\item\label{i06}
any configuration
obtained by flipping to zero the spins
associated with any nearest neighbor connected
subset of the set of sites adjacent to either the horizontal or vertical
sides of any rectangular droplet $\zeta\in\mathcal{R}_{\ell_1,\ell_2}$
such that
$\ell_1=\ell_2=L-2$;
\item\label{i07}
any configuration
obtained by adding an $\ell$ protuberance to one of the
two sides of the stripe
$\zeta\in\mathcal{R}_{\ell_1,\ell_2}$
such that
$\ell_1=L$ and $\ell_\textrm{c}\le\ell_2\le L-3$;
\item\label{i08}
any configuration different from $\zero$
obtained by flipping to zero the spins
associated with any nearest neighbor connected
subset of the set of sites associated with minus one spis
of any rectangular droplet $\zeta\in\mathcal{R}_{\ell_1,\ell_2}$
such that
$\ell_1=L$ and $\ell_2=L-2$;
\item\label{i09}
all the configurations in $\mathcal{P'}_c$, namely
those configuration obtained by adding a $0$ protuberance of length one
the side long $\ell_c -1$ of any rectangle
$\zeta\in\mathcal{R}_{\ell_\textrm{c}-1,\ell_\textrm{c}}$;
\item\label{i10}
the non--trivial cycles
$\mathcal{A}_\zeta$ for $\zeta\in  \mathcal{R}_{\ell_c -1,\ell_c +1}$.
\end{enumerate}

The $\mathcal{T}_\zero(\muno)$
satisfies the following properties:
$\zero \notin  \mathcal{T}_{\zero}(\muno)$,
$\puno \notin  \partial\mathcal{T}_{\zero}(\muno)
\cup \mathcal{T}_{\zero}(\muno) $,
there exists a cycle $C\in\mathcal{T}_\zero(\muno)$
such that $\zero\in \cc{B}(C)$,
and for any cycle $C\in\mathcal{T}_\zero(\muno)$
\begin{equation}
\label{dentro}
\mathcal{B}(C)
\subset
\bigcup_{D\in\mathcal{T}_\zero(\muno)}D
\cup\{\zero\}
\end{equation}
where it is useful to emphasize that $\mathcal{T}_\zero(\muno)$ is a
collection of pairwise disjoint cycles.

The first two properties are immediate by definition. We comment briefly on
the last two.
At step~\ref{i06} in the definition of $\mathcal{T}_\zero(\muno)$
we added in particular the configurations in which all the spins on
the lattice are equal to zero but one which is equal to minus one.
It is immediate to realize that the principal boundary of such a
trivial cycle is the set $\{\zero\}$.
Finally, to prove equation \eqref{dentro} one has to examine all
the cycles included in $\mathcal{T}_\zero(\muno)$ and
prove that their principal boundaries are subset of the
right hand side of equation \eqref{dentro}. With the same labelling used
in the costruction of the set $\mathcal{T}_\zero(\muno)$ we have:
\ref{i01}.\ by the methods of proof
of \cite[Lemma~4.12]{CN2013}
we have that $\mathcal{B}(\mathcal{D}_\zero(\muno))=\pcri\cup\pcri'$.
\ref{i02}.\
for any $\zeta\in\mathcal{P}_\textrm{c}$
the set $\mathcal{B}(\zeta)$
is made of the configurations obtained by enlarging by one site
the protuberance; those configurations
belong to the not trivial cycles added at step \ref{i03}.
\ref{i03}.\
from item $i)$ in Lemma~\ref{l:rett0} it follows that the principal
boundary of the cycle $\mathcal{A}_\zeta$ is made by the configurations
obtained by adding a protuberance to one of the four sides of the
rectangle $\zeta$; these configurations are added at steps \ref{i04}
and \ref{i05};
\ref{i04}.\ consider the configuration $\zeta$ and
suppose the protuberance is on the side of length $\ell_1$
(the argument is analogous in the other case), then
$\mathcal{B}(\zeta)$
is made of the configurations obtained by enlarging by one site
the protuberance; those configurations are either trivial cycles
added to $\mathcal{T}_\zero(\muno)$ at step \ref{i04}
or elements of the not trivial cycles added at step \ref{i03}.
\ref{i05}.\
consider the configuration $\zeta$ and assume that it is not a stripe
wind around the torus, then
$\mathcal{B}(\zeta)$
is made of the configurations obtained by flipping to zero any
minus spin with precisely two neighboring zeros;
those configurations are
added to $\mathcal{T}_\zero(\muno)$ at step \ref{i05}.
On the other hand, if $\zeta$ is a stripe winding around the torus
the principal boundary is the set of configurations obtained by adding
a zero protuberance; those configurations are added to
$\mathcal{T}_\zero(\muno)$ at step \ref{i07}.
The discussion of the other cases is similar.
We just discuss explicitly the case \ref{i10}:
by Lemma~\ref{l:rett1} we have that
the principal
boundary of the cycle $\mathcal{A}_\zeta$ is made of the configurations
obtained by flipping to minus  all the spins but one spin of one of
the two shortest sides of the rectangular droplet $\zeta$;
those configurations are elements of $\pcri^\prime$.

\begin{lemma}
\label{l:exit}
Consider the set  $\mathcal{T}_{\zero}(\muno)$. Then there exists $\kappa>0$ such that for $\beta$ sufficiently large:
$$
\bb{P}_\muno(\tau_{\partial \mathcal{T}_{\zero}(\muno)} < \tau_\zero)\leq e^{-\beta\kappa}
$$
\end{lemma}
\par\noindent
\textit{Proof.\/}
We have that
\begin{eqnarray*}
\mathbb{P}_\muno(\tau_{\partial \mathcal{T}_{\zero}(\muno)} < \tau_{\zero})&=&\hspace{-0.4cm}\sum_{C\in  \mathcal{T}_{\zero}(\muno)}
\mathbb{P}_\muno(\tau_{\partial \mathcal{T}_{\zero}(\muno)} <\tau_{\zero}, X_{\tau_{\partial \mathcal{T}_{\zero(\muno)}-1}}\in C,X_{\tau_{\partial \mathcal{T}_{\zero(\muno)}}}\notin \mathcal{B}(C))\\
&=&\hspace{-0.3cm}\sum_{C\in  \mathcal{T}_{\zero}(\muno)}\sum_{z\in C}
\mathbb{P}_\muno(\tau_{\partial \mathcal{T}_{\zero}(\muno)} < \tau_{\zero}, X_{\tau_{\partial \mathcal{T}_{\zero(\muno)}-1}}=z,X_{\tau_{\partial \mathcal{T}_{\zero(\muno)}}}\hspace{-0.3cm}\notin  \mathcal{B}(C))\\
&\leq& \hspace{-0.3cm}\sum_{C\in  \mathcal{T}_{\zero}(\muno)}  \sum_{z\in C}
\mathbb{P}_z(X_{\tau_{\partial C}}\notin  \mathcal{B}(C))
\leq \sum_{C\in  \mathcal{T}_{\zero}(\muno)} \vert C\vert
e^{- K_C \beta}<e^{-\kappa\beta}
\end{eqnarray*}
where in the first equality we have used the fact that  for each cycle  $C$ in $\mathcal{T}_{\zero}(\muno)$  the principal boundary $\mathcal{B}(C)$  is contained in $\mathcal{T}_{\zero}(\muno)$; the first inequality follows from the strong Markov property and
the second inequality is a consequence of \cite[Theorem 6.23]{OV}.
\qed

\medskip
\par\noindent
\textit{Proof of Lemma~\ref{t:bc020}.\/}
Since $\muno\in \mathcal{T}_{\zero}(\muno)$, $\zero\in \partial \mathcal{T}_{\zero}(\muno)$ and $\puno\notin \mathcal{T}_{\zero}(\muno)\cup \partial \mathcal{T}_{\zero}(\muno)$,
we have that,
$$
\mathbb{P}_\muno(\tau_\puno<\tau_\zero)\leq \mathbb{P}_\muno(\tau_{\partial\mathcal{T}_\zero(\muno)}<\tau_\zero)
\;\;.
$$
The lemma then follows from Lemma~\ref{l:exit}.
\qed


\medskip
\par\noindent
\textit{Proof of Lemma~\ref{t:bc040}.\/}
The proof of this lemma could be given by applying Proposition 2.3  and equation (5.1) in \cite{LL2016}, correcting by a factor $2|\Lambda|$ for passing from the continuos to the discrete time  of the  Metropolis dynamics. For the sake of completeness, we provide an alternative proof, based  on Theorem~\ref{t:6.19} contained in Appendix~\ref{s:appendix:B}.
By \cite[Lemma~4.9]{CN2013}, we have that
$\Phi(\muno,\{\zero,\puno\})=\Phi(\muno,\puno)=
\Gamma_\textrm{c}+H(\muno)$.
Moreover, by \cite[Theorem~4.14]{CN2013}, we know that
the minimal gates
$\gate$ between $\muno$ and $\puno$
is $\cc{P}_\textrm{c}$. Hence, by Theorem~\ref{t:6.19},
\begin{equation}
\label{emi000}
\rr{cap}_\beta(\muno,\{\zero,\puno\})
=\frac{k_1}{Z_\beta} e^{-\beta\Phi(\muno,\{\zero,\puno\})}[1+o(1)]
\end{equation}
with
\begin{equation}
\label{K1}
k_1=\sum_{z\in\cc{P}_\textrm{c}}\frac{\check{p}(z)\hat{p}(z)}{\check{p}(z)+\hat{p}(z)}
\end{equation}
with
$\check{p}(z)=1/(2|\Lambda|)$,
since the only possible transition to $\cc{W}$ is removing the protuberance of the protocritical droplet.
On the other hand,
$\hat{p}(z)=1/(2|\Lambda|)$
if the protuberance is on the corner and
$\hat{p}(z)=1/|\Lambda|$ otherwise. So that,
$$
k_1=\frac{1}{2|\Lambda|}{4|\Lambda|}\left(2\times\frac{1}{2}
+(\ell_\textrm{c}-2)\frac{2}{3}\right)=
\frac{2}{3}\left(1+(\ell_\textrm{c}-2)\frac{2}{3}\right)
=\frac{2}{3}(2\ell_\textrm{c}-1)
\;\;.
 $$
Therefore,
\begin{displaymath}
\rr{cap}_\beta(\muno,\{\zero,\puno\})=\frac{2}{3}(2\ell_\textrm{c}-1)\frac{e^{-\beta(\Gamma_\textrm{c}+H(\muno))}}{Z_\beta}
[1+o(1)]
\;\;.
\end{displaymath}

As regards the calculation of $\rr{cap}_\beta(\zero,\puno)$,
by \cite[Lemma~4.9]{CN2013}, we have that $\Phi(\zero,\puno\})
=\Gamma_\textrm{c}+H(\zero)$. Moreover,
by \cite[Theorem~4.14]{CN2013}, we know that
the union of the minimal gates between $\zero$ and $\puno$ is
${Q}_\textrm{c}$. Hence,
\begin{displaymath}
\rr{cap}_\beta(\zero,\puno)
=\frac{k_2}{Z_\beta} e^{-\beta\Phi(\zero,\puno)}[1+o(1)]
\end{displaymath}
with
\begin{displaymath}
k_2=\sum_{z\in\cc{Q}_\textrm{c}}\frac{\check{p}(z)\hat{p}(z)}{\check{p}(z)+\hat{p}(z)}
\end{displaymath}
With similar arguments as above,
we get $k_2=k_1$
and, therefore,
\begin{displaymath}
\rr{cap}_\beta(\zero,\puno)=\frac{2}{3}(2\ell_\textrm{c}-1)\frac{e^{-\beta(\Gamma_\textrm{c}+H(\zero))}}
{Z_\beta}
\end{displaymath}
which completes the proof of the lemma.
\qed

\medskip

\noindent\textbf{Aknowledgements.} ENMC thanks ICMS (TU/e, Eindhoven), Eurandom (TU/e, Eindhoven), the Mathematics Department of Delft University, and the Mathematics Department of Utrecht University for kind hospitality. FRN and CS thank A. Bovier for many stimulating discussions. ENMC and FRN thank E. Scoppola and F. den Hollander for illuminating discussions. The authors thank A. Gaudilliere and M. Slowik for useful discussions and comments and an anonymous referee, whose  useful comments  considerably improved the quality of the manuscript.
\appendix
\section{General bounds}
 \setcounter{equation}{0} 
     \setcounter{theorem}{0} 
\label{s:generali}
\par\noindent
In this appendix we summarize some general results whose statement
and proof can already
be found in the literature but, sometimes, in slightly different contexts.

\begin{proposition}{\normalfont\textrm{($\!\!$\cite[Lemma 3.1.1]{BHN})}}
\label{t:apriori}
Consider the Markov chain defined in Section~\ref{s:reversibile}.
For every not empty disjoint sets $Y,Z\subset X$ there exist
constants $0<C_1<C_2<\infty$ such that
\begin{equation}
\label{r:apriori}
C_1\le e^{\beta\Phi(Y,Z)}\, Z_\beta\, \rr{cap}_\beta(Y,Z)\le C_2
\end{equation}
for all $\beta$ large enough.
\end{proposition}

\medskip
\par\noindent
\textit{Proof.\/}
The upper bound can be obtained by choosing
$f=\mathbb{I}_{K(Y,Z)}$
in \eqref{capac}
with
$$
K(Y,Z):=\{x\in X\setminus Y:\Phi(x,Y)\le\Phi(Y,Z)\}
$$
For any pair $u,v\in X$ such that
$u\in K(Y,Z)$ and $v\in X\setminus K(Y,Z)$, we have that
$H(u)+\Delta(u,v)\geq \Phi(Y,Z)$. In fact, if by absurdity
it were
$H(u)+\Delta(u,v)<\Phi(Y,Z)$, it would be possible to construct a
path $\omega$ starting at $v$ and ending in $Y$ such that
$\Phi_\omega< \Phi(Y,Z)$, which is
in contradiction with $v\in X\setminus K(Y,Z)$.
Hence, by (\ref{diri-new}) and (\ref{capac})
$$
Z_\beta
\rr{cap}_\beta(Y,Z)\leq
Z_\beta
\mathscr{D}_\beta[\mathbb{I}_{K(Y)}]
=
\frac{1}{2}
\sum_{{u\in K(Y,Z)}\atop{v\in X\setminus K(Y,Z)}}\,
p_\beta(u,v)
  e^{-G_\beta(u)}
$$
Recalling \eqref{rev04} and \eqref{rev04-gib},
we get that there exists $C$ such that for $\beta$
large enough
$$
Z_\beta
\rr{cap}_\beta(Y,Z)
\le
\sum_{{u\in K(Y)}\atop{v\in X\setminus K(Y)}}\,
 C
  e^{-\beta[H(u)+\Delta(u,v)]}
$$
Finally, the upper bound in \eqref{r:apriori} follows from the
fact that
$H(u)+\Delta(u,v)\geq \Phi(Y,Z)$
for any
$u\in K(Y,Z)$ and $v\in X\setminus K(Y,Z)$.

As regards the lower bound, it can be obtained by picking any path
$\omega=(\omega_0,\omega_1,\ldots,\omega_n)$
that realizes the minimax in $\Phi(Y,Z)$ and ignore
all the transitions that are not the path and using the same argument as in the proof of  [Lemma~3.1.1, \cite{BHN}].
An alternative proof can be given applying the Berman--Konsowa lemma [Proposition 2.4, \cite{BHS}] which provides a complementary variational principle, in the sense that any test flow will give a lower bound.
Hence, the lower bound can be obtained by picking any path
$\omega=(\omega_0,\omega_1,\ldots,\omega_n)$,
with $\omega_0\in Y$,  $\omega_n\in Z$, such that it realizes the minimax in $\Phi(Y,Z)$  and such that
$H(\omega_i)+ \Delta(\omega_i,\omega_{i+1})\leq \Phi(Y,Z)$ for $i\in\{0,\ldots,n-1\}$ (recall (\ref{height}) and (\ref{communication-set})). If we choose a unitary flow  for the edges in the path $\omega$ and null otherwise, the induced Markov chains is a deterministic chain along the path, so that the expectation in [Proposition 2.4, \cite{BHS}]  is just the contribution of the deterministic path. Hence, for the chosen flow we have:
$$
\rr{cap}_\beta(Y,Z)\geq \left[\sum_{k=0}^{n-1}\frac{1}{\mu_\beta(\omega_k) p_\beta(\omega_k,\omega_{k+1})} \right]^{-1}\geq C_1\frac{1}{Z_\beta}
e^{-\beta\Phi(Y,Z)}[1+o(1)]
$$
where in the last inequality we used
\eqref{rev04} and \eqref{rev04-gib}.
\qed

\begin{proposition}
\label{t:cri02}
Consider the Markov chain defined in Section~\ref{s:reversibile}.
We have that
\begin{equation}
\label{potbound}
\mathbb{P}_y(\tau_{y_1}<\tau_{y_2})\leq \frac{\rr{cap}_\beta(y,y_1)}{\rr{cap}_\beta(y,y_2)}
\end{equation}
for any $y\neq y_1$, $y_1 \neq y_2$, $y\neq y_2$, and $y,y_1,y_2\in X$.
\end{proposition}

\medskip
\par\noindent
\textit{Proof.\/}
Given $y,y_1,y_2\in X$, a renewal argument and the strong Markov property yield:
\begin{eqnarray*}
\mathbb{P}_y(\tau_{y_1}<\tau_{y_2})&=&\mathbb{P}_y(\tau_{y_1}<\tau_{y_2}, \tau_{\{y_1,y_2\}}>\tau_{y})+\mathbb{P}_y(\tau_{y_1}<\tau_{y_2}, \tau_{\{y_1,y_2\}}<\tau_{y})\\
&=&\mathbb{P}_y(\tau_{y_1}<\tau_{y_2}| \tau_{\{y_1,y_2\}}>\tau_{y})\mathbb{P}_y(\tau_{\{y_1,y_2\}}> \tau_{y})\\
&&+\mathbb{P}_y(\tau_{y_1}<\tau_{y_2}, \tau_{\{y_1,y_2\}}<\tau_{y})\\
&=&\mathbb{P}_y(\tau_{y_1}<\tau_{y_2})\mathbb{P}_y(\tau_{\{y_1,y_2\}}> \tau_{y})+\mathbb{P}_y(\tau_{y_1}<\tau_{y_2}, \tau_{y_1}<\tau_{y})\\
&=&\mathbb{P}_y(\tau_{y_1}<\tau_{y_2})\mathbb{P}_y(\tau_{\{y_1,y_2\}}> \tau_{y})+\mathbb{P}_y(\tau_{y_1}<\tau_{\{y_2,y\}})
\end{eqnarray*}
Therefore
\begin{displaymath}
\mathbb{P}_y(\tau_{y_1}<\tau_{y_2})
\!=\!
\frac{\mathbb{P}_y(\tau_{y_1}<\tau_{\{y_2, y\}})}
{1-\mathbb{P}_y(\tau_{\{y_1,y_2\}}>\tau_{y})}=
\frac{\mathbb{P}_y(\tau_{y_1}<\tau_{\{y_2, y\}})}
{\mathbb{P}_y(\tau_{\{y_1,y_2\}}< \tau_{y})}
\leq \frac{\mathbb{P}_y(\tau_{y_1}<\tau_{y})}
{\mathbb{P}_y(\tau_{ y_2} <\tau_{y})}
\end{displaymath}
Recalling \eqref{cap-prop},
we can rewrite the ratio in terms of ratio of capacities:
\begin{displaymath}
\label{cri02:023}
\frac{\mathbb{P}_y(\tau_{y_1}<\tau_{y})}
{\mathbb{P}_y(\tau_{ y_2} <\tau_{y})}=\frac{\rr{cap}_\beta(y,y_1)}{\rr{cap}_\beta(y,y_2)}
\end{displaymath}
Hence, we get \eqref{potbound}.
\qed

\section{Capacity estimate for Metropolis dynamics}
\label{s:appendix:B}
\setcounter{equation}{0} 
     \setcounter{theorem}{0} 
\par\noindent
In this section we state and prove  a slightly  more 
general theorem than Theorem~6.19 of \cite{MNOS}.
As in  \cite{MNOS}, the theorem holds for the Metropolis
dynamics introduced in Section~\ref{s:esempi}, but in 
the more general setting of two metastable configurations. 
We assume that Condition~\ref{t:series00} holds
so that the energy landscape is such that
there exist three states $\metadue$, $\metauno$, and $\stab$
such that
$X_\rr{s}=\{\stab\}$,
$X_\rr{m}=\{\metauno,\metadue\}$,
and
$K(\metadue)>K(\metauno)$.
This theorem gives indeed
the capacity between a configuration $x\in \{ \metadue,\metauno\}$ 
and a set $A=\{ \stab,\metauno\}\setminus \{x\}$ in terms
of the  energy and cardinality of the minimal gates.
Hence, the theorem considers the two case  
$x=\metadue$, $A=\{\metauno,\stab\}$ and  $x=\metauno$, $A=\{\stab\}$.

For $x$ and $A$ as above we let 
$Q_x:=\{y\in X: \Phi(y,x)<\Phi(x,A)\}$ and
$Q_A:=\{y\in X: \Phi(y,A)<\Phi(x,A)\}$. 
By Definition~\ref{def1} and Condition~\ref{t:series00} 
we have that $Q_A$ is a cycle and 
$Q_A\cap Q_x=\emptyset$. 

Before stating the theorem we recall first some notions
introduced in \cite{MNOS}. Let $z\in X$ and $B\subset X$. 
We say that $\mathcal{S}(z,B)\subset X$  is the 
\definisco{set of saddles} for the pair $z$ and $B$
if and only if $K(z)=\Phi(z,B)$
for any $z\in \saddle(z,B)$.
We say that a set $\gate\subseteq\saddle(z,B)$ is a
\definisco{gate} for the pair $z$ and $B$ if 
for any path $\omega\in\Omega(z,B)$ such that
$\Phi_\omega=\Phi(z,B)$ we have that  $\gate\cap \omega\neq \emptyset$.
A gate $\gate$ for the pair $z$ and $B$
is said to be \definisco{minimal} if and only if for any
proper subset $\gate^\prime$ of $\gate$ there exists
a path joining $z$ to $B$ with maximal height equal to the
communication height between $z$ and $B$
which does not pass through $\gate^\prime$.

\begin{theorem}
\label{t:6.19}
In the setup introduced above, consider
$x\in \{\metadue,\metauno\}$ and 
$A=\{ \stab,\metauno\}\setminus \{x\}$.
Assume that the minimal gate $\gate$ for $x$ and $A$ is unique
and that 
for any $y\in\gate$ and $w\notin Q_x\cup Q_A$ we have $p(y,w)=o(1)$.
Then
\begin{equation}
\label{cap_gate}
\rr{cap}_\beta(x,A)=\frac{k}{Z_\beta} e^{-\beta\Phi(x,A)}[1+o(1)]
\end{equation}
with
\begin{displaymath}
k=\sum_{z\in \gate}\frac{\check{p}(z)\hat{p}(z)}{\check{p}(z)+\hat{p}(z)}
\end{displaymath}
where $\check{p}(z):=\sum_{w\in Q_x}p_\beta(z,w)$ and
$\hat{p}(z):=\sum_{w\in Q_A}p_\beta(z,w)$.
\end{theorem}
\medskip
\par\noindent
\textit{Proof.\/}
Upper bound:
by (\ref{cap-prop}), \cite[Theorem 5.4]{MNOS} and using the strong Markov property, we can write:
 \begin{equation}
 \label{cap_split_1}
 \rr{cap}_\beta(x,A)
=\mu_{\beta}(x)\sum_{z\in \gate}
{\bb{P}}_x(\tau_z<\tau_{\gate\setminus z}, \tau_z<\tau_x) 
  {\bb{P}}_z(\tau_{A}< \tau_x)[1+o(1)]
 \end{equation}
 In fact we have
 \begin{eqnarray*}
 \rr{cap}_\beta(x,A)&=&\mu_{\beta}(x)
{\bb{P}}_x( \tau_A<\tau_x) =\mu_{\beta}(x)
{\bb{P}}_x( \tau_A<\tau_x,\tau_{\gate}<\tau_A)[1+o(1)]\\
&=&\mu_{\beta}(x)
\sum_{z\in\gate}{\bb{P}}_x( \tau_A<\tau_x,\tau_{z}<\tau_A,\tau_z<\tau_{\gate\setminus z})[1+o(1)]\\
&=&\mu_{\beta}(x)
\sum_{z\in\gate}{\bb{P}}_x( \tau_A<\tau_x|\tau_{z}<\tau_A,\tau_z<\tau_{\gate\setminus z},\tau_z<\tau_x)\\
&&
\phantom{\mu_{\beta}(x)\sum_{z\in\gate}}
\times{\bb{P}}_x(\tau_{z}<\tau_A,\tau_z<\tau_{\gate\setminus z},\tau_z<\tau_x)[1+o(1)]\\
&=&\mu_{\beta}(x)\sum_{z\in\gate}{\bb{P}}_z( \tau_A<\tau_x) {\bb{P}}_x(\tau_{z}<\tau_A,\tau_z<\tau_{\gate\setminus z},\tau_z<\tau_x)[1+o(1)]\\
&=&\mu_{\beta}(x)\sum_{z\in\gate}{\bb{P}}_z( \tau_A<\tau_x) {\bb{P}}_x(\tau_z<\tau_{\gate\setminus z},\tau_z<\tau_x)[1+o(1)]
 \end{eqnarray*}
 where in the last step we used the fact that:
 $${\bb{P}}_x(\tau_{z}<\tau_A,\tau_z<\tau_{\gate\setminus z})=
 {\bb{P}}_x(\tau_{\gate}<\tau_A,\tau_z<\tau_{\gate\setminus z})= {\bb{P}}_x(\tau_z<\tau_{\gate\setminus z})[1+o(1)]$$
For the second term of \ref{cap_split_1}, we have, for any $z\in \gate$
\begin{eqnarray}
\bb{P}_z(\tau_A<\tau_x)
&=&\Big[\sum_{y\in Q_A} p(z,y) \bb{P}_y(\tau_A<\tau_x)
+\sum_{y\in Q_x} p(z,y) \bb{P}_y(\tau_A<\tau_x) \nonumber \\
&&
\phantom{\Big[}
+p(z,z) \bb{P}_z(\tau_A<\tau_x)\Big][1+o(1)] \nonumber \\
&=&\frac{1}{1-p(z,z)}(\sum_{y\in Q_A} p(z,y) \bb{P}_y(\tau_A<\tau_x) 
+\!\!\sum_{y\in Q_x} p(z,y) \bb{P}_y(\tau_A<\tau_x))[1+o(1)] \nonumber\\
&\leq &\frac{\hat{p}(z)}{\hat{p}(z)+\check{p}(z)}[1+o(1)] \label{e:pa}
\end{eqnarray}
where in the first equality we used the fact that   $p(z,y)=o(1)$ 
 for any $y\in (Q_x\cup Q_A\cup\{z\})^c$;  in the last step, 
 we  have used the trivial relation $\bb{P}_y(\tau_{A}<\tau_{x})\leq1$  for bounding the first term of the sum, while the upper bound for the second term follows from
 $ \bb{P}_y(\tau_{A}<\tau_{x})\leq \bb{P}_y(\tau_{\partial Q_x}<\tau_{x})$ and from  the recurrence of non trivial cycle (i.e., by Theorem~6.3 in \cite{OV},
 for any $y\in Q_x$ we have
 $ \bb{P}_y(\tau_{\partial Q_x}<\tau_{x})=o(1)$). 
Hence, by (\ref{e:pa}) and (\ref{cap_split_1}) we have:
\begin{eqnarray}
\rr{cap}_\beta(x,A)
&\leq&\mu_{\beta}(x)\sum_{z\in \gate} \frac{\hat{p}(z)}{\hat{p}(z)+\check{p}(z)}
{\bb{P}}_x(\tau_z<\tau_{\gate\setminus z}, \tau_z<\tau_x)\nonumber  [1+o(1)]  \\
&=& \mu_{\beta}(x)\sum_{z\in \gate} \frac{\hat{p}(z)}{\hat{p}(z)+\check{p}(z)}
\frac{\mu_\beta(z)}{\mu_\beta(x)}{\bb{P}}_z(\tau_x<\tau_\gate)  [1+o(1)] \nonumber\\
&\leq&\sum_{z\in \gate}  \mu_{\beta}(z) \frac{\hat{p}(z)}{\hat{p}(z)+\check{p}(z)}
\sum_{y\in Q_x}p(z,y) [1+o(1)]
\leq\frac{e^{-\beta \Phi(x,A)}}{Z_\beta}\sum_{z\in \gate}  \frac{\hat{p}(z)\check{p}(z)}{\hat{p}(z)+\check{p}(z)}\nonumber
  \end{eqnarray}
 where in the second step we used reversibility and in the third one: 
 $$\bb{P}_z(\tau_x<\tau_z)=\Big[\sum_{y\in Q_A} p(z,y) \bb{P}_y(\tau_x<\tau_\gate)+\sum_{y\in Q_x} p(z,y) \bb{P}_y(\tau_x<\tau_\gate)\Big][1+o(1)]$$
  and the properties of the cycles $Q_x$ and $Q_A$.

\emph{Lower bound}.\
In order to prove the lower bound, we adapt to our setting the arguments of the proof in \cite[Lemma 3.2]{BM}.
We consider a subgraph $\Delta$ in the space of configurations obtained
removing all the connections to the  configurations in $\partial Q_x\setminus \gate$.
We denote with
$\widetilde{\bb{P}}_x$ the probability along the trajectories of this
\emph{restricted} process started at $x$. Hence, by Rayleigh's
shortcut rule (see \cite[Lemma 4.2]{BM}) we have:
 \begin{displaymath}
 {\rr{cap}}_\beta(x,A)\ge \widetilde{\rr{cap}}_\beta(x,A)
 \end{displaymath}
If we now exploit the property of the restricted process we can write:
 \begin{equation}
 \label{cap_split}
 \widetilde{\rr{cap}}_\beta(x,A)
=\mu_{\beta}(x)\sum_{z\in \gate}
\widetilde{\bb{P}}_x(\tau_z<\tau_{\gate\setminus z}, \tau_z<\tau_x)
  \widetilde{\bb{P}}_z(\tau_{A}< \tau_x)
 \end{equation}
where we used (\ref{cap-prop}) and the strong Markov property.
By reversibility, (\ref{cap_split}) becomes
\begin{eqnarray}
\widetilde{\rr{cap}}_\beta(x,A)&=&
\mu_\beta(x)\sum_{z\in \gate}
\frac{\mu_\beta(z)}{\mu_\beta(x)}\widetilde{\bb{P}}_z(\tau_x< \tau_{ \gate})
  \widetilde{\bb{P}}_z(\tau_{A}< \tau_x) \nonumber \\
  &=&\frac{e^{-\beta\Phi(x,A)}}{Z_\beta} \sum_{z\in \gate}
\widetilde{\bb{P}}_z(\tau_x< \tau_{\gate})
  \widetilde{\bb{P}}_z(\tau_{A}< \tau_x)
 \label{captilde}
 \end{eqnarray}
Now, the first factor in (\ref{captilde}) can be rewritten as
 \begin{displaymath}
 \widetilde{\bb{P}}_z(\tau_x< \tau_{\gate})\geq
\sum_{y\in Q_x} \tilde{p}_\beta(z,y)
\widetilde{\bb{P}}_y(\tau_x< \tau_{ \gate})=
  \sum_{y\in Q_x} \tilde{p}_\beta(z,y)\tilde{h}_{x, \gate}(y)
 \end{displaymath}
We want to prove that inside $Q_x$
the equilibrium potential $\tilde{h}_{x,\gate}(y)$
is exponentially close to 1. Therefore,
for any $y \in Q_x $, and for $\delta=\Phi(x,A)-K(y)>0$:
\begin{displaymath}
1-\tilde{h}_{x,\gate}(y)=\tilde{h}_{\gate,x}(y)\leq
\frac{\widetilde{\rr{cap}}_\beta(y,\gate)}{\widetilde{\rr{cap}}_\beta(y,x)}
\leq C\frac{e^{-\beta \Phi(y,\gate)}}{e^{-\beta \Phi(y,x)}}
=C\,\frac{e^{-\beta \Phi(x,A)}}{e^{-\beta (\Phi(x,A)-\delta)}}
=C\,e^{-\beta\delta}
\end{displaymath}
where in the first inequality we used Proposition~\ref{t:cri02}, and
in the second inequality Proposition~\ref{t:apriori}.
Therefore
\begin{equation}
\label{f1}
\widetilde{\bb{P}}_z(\tau_x< \tau_{\gate})\geq
\sum_{y\in Q_x} \tilde{p}_\beta(z,y) [1+o(1)]= \check{p}(z) [1+o(1)]
\end{equation}
where in the last step, we used the fact that the one step
transition probabilities $\tilde{p}(z,\cdot)$ are equal,
by construction, to ${p}(z,\cdot)$  of the original chain,
for any $z\in \gate$.
For the second factor $ \widetilde{\bb{P}}_z(\tau_A<\tau_x)$:
\begin{equation*}
\widetilde{\bb{P}}_z(\tau_A<\tau_x)\geq
\tilde{p}_\beta(z,z) \widetilde{\bb{P}}_z(\tau_A<\tau_x)
+\sum_{y\in {Q}_A} \tilde{p}_\beta(z,y)\widetilde{\bb{P}}_y(\tau_A< \tau_x)
 \end{equation*}
 so that
 \begin{displaymath}
\widetilde{\bb{P}}_z(\tau_A<\tau_x)\ge
\frac{1}{1- \tilde{p}_\beta(z,z)}
\sum_{y\in {Q}_A} \tilde{p}_\beta(z,y)\tilde{h}_{A,x}(y)
=\frac{1}{1- {p}_\beta(z,z)}
\hspace{-0.2cm} \sum_{y\in {Q}_A} {p}_\beta(z,y)\tilde{h}_{A,x}(y)
 \end{displaymath}
We want to prove that inside ${Q}_A$ the equilibrium potential
$\tilde{h}_{A,x}(y)$ is exponentially close to 1. Therefore for any $y \in {Q}_A$, by using Proposition~\ref{t:cri02}, and
 Proposition~\ref{t:apriori} we have
\begin{displaymath}
1-\tilde{h}_{A,x}(y)=\tilde{h}_{x,A}(y)\leq
\frac{\widetilde{\rr{cap}}_\beta(y,x)}{\widetilde{\rr{cap}}_\beta(y,A)}
\leq C\frac{e^{-\beta \Phi(y,x})}{e^{-\beta \Phi(y,A)}}
=C\,\frac{e^{-\beta \Phi(x,A)}}{e^{-\beta (\Phi(x,A)-\delta)}}
=C\,e^{-\beta\delta}
\end{displaymath}
so that
\begin{equation}
 \label{factor22}
\widetilde{\bb{P}}_z(\tau_A<\tau_x)\geq\frac{1}{\hat{p}(z)+\check{p}(z)}
  \sum_{y\in {Q}_A} {p}_\beta(z,y) [1+o(1)]
=\frac{\hat{p}(z)}{{\hat{p}(z)+\check{p}(z)}}[1+o(1)]
   \end{equation}
Hence, by (\ref{captilde}), (\ref{f1}),
and (\ref{factor22}),  (\ref{cap_gate}) follows.
\qed


\end{document}